\documentclass[aps,prb,reprint,floatfix,groupedaddress,superscriptaddress,showpacs]{revtex4-1}

\usepackage{lipsum}
\usepackage{graphicx}
\usepackage{natbib}
\usepackage[utf8]{inputenc}
\usepackage{tikz}
\usepackage{amsmath}
\usepackage{empheq}
\usepackage{geometry}
\usepackage{amssymb}
\usepackage{color}
\usepackage{multirow}
\usepackage{hyperref}

\begin{document}
  
\title{Linear spin wave theory for single-Q incommensurate magnetic structures}
\author{S. Toth}
\email{sandor.toth@psi.ch}
\affiliation{Laboratory for Neutron Scattering, Paul Scherrer Institut (PSI), CH-5232 Villigen, Switzerland}
\affiliation{Helmholtz-Zentrum Berlin, Hahn-Meitner Platz 1, D-14109 Berlin, Germany}
\affiliation{Laboratory for Quantum Magnetism, ICMP, Ecole Polytechnique Fédérale de Lausanne (EPFL), CH-1015 Lausanne, Switzerland}
\author{B. Lake}
\affiliation{Helmholtz-Zentrum Berlin, Hahn-Meitner Platz 1, D-14109 Berlin, Germany}
\affiliation{Institut für Festkörperphysik, Technische Universität Berlin, Hardenbergstraße 36, D-10623 Berlin, Germany}
\date{\textrm{\today}}
\pacs{75.10.Hk, 75.30.Ds, 75.30.Et}
% 75.10.Hk Classical spin models
% 75.30.Ds Spin waves
% 75.30.Et Exchange and superexchange interactions

\newcommand{\vect}[1]{{\bf{#1}}}
\newcommand{\matr}[1]{\mathsf{#1}}
\newcommand{\acacro}{$\alpha$-CaCr$_2$O$_4$}
\newcommand{\ede}{\epsilon_\Delta}
\newcommand{\ehe}{\epsilon_H}
\newcommand{\ete}{\epsilon_T}

\begin{abstract}
  Linear spin wave theory provides the leading term in the calculation of the excitation spectra of long-range ordered magnetic systems as a function of $1/\sqrt{S}$. This term is acquired using the Holstein-Primakoff approximation of the spin operator and valid for small $\delta S$ fluctuations of the ordered moment. We propose an algorithm that allows magnetic ground states with general moment directions and single-Q incommensurate ordering wave vector using a local coordinate transformation for every spin and a rotating coordinate transformation for the incommensurability. Finally we show, how our model can determine the spin wave spectrum of the magnetic C-site langasites with incommensurate order.
\end{abstract}

\maketitle

\section{Introduction}

  Linear spin wave theory (LSWT) was first introduced by Bloch [\onlinecite{Bloch1930a}] and independently by Slater [\onlinecite{Slater1930}]. The description using second quantization of bosonic operators was developed by Holstein and Primakoff [\onlinecite{Holstein1940a}] with subsequent theoretical development by Dyson [\onlinecite{Dyson1956b}, \onlinecite{Dyson1956}] to described spin-wave interactions. The concept of spin waves was a milestone in understanding the magnetic correlations in ordered systems. However after decades, the focus was moved onto new areas in magnetism as new theory and materials were developed. One of the main area of recent interest is frustrated magnetism. Frustration leads to exciting novel states of matter such as spin ice \cite{Fennell2009, Morris2009}, spin liquid \cite{Balents2010} and multiferroic phases \cite{Khomskii2009a}. The competing nature of the interactions often leads to non-collinear magnetic structures with incommensurate order. To identify possible exchange pathways and energies in these materials, modeling the magnetic excitation spectrum is essential. Linear spin wave theory combined with neutron and high-resolution resonant inelastic X-ray scattering \cite{Kim2012a} provide a powerful toolset to understand the magnetic interactions in these materials in full details.

  The spin wave excitations of long range ordered magnetic systems are well understood, however dealing with large magnetic unit cells, several competing spin-spin interactions and incommensurate magnetic order are still challenging due to the lack of a general algorithm. S.\ Petit [\onlinecite{Petit2011}] and J. Haraldsen et al.\ [\onlinecite{Haraldsen2010b}] both developed a general spin wave theory, limited to commensurate magnetic structures with canted spins and isotropic exchange interactions. Their formalism can be applied with limitations to incommensurate order, by extending the magnetic unit cell to approximate the incommensurate magnetic ordering wave vector with a rational number. Here we propose an extension to his method, where the magnetic ordering wave vector can be arbitrary. The proposed method gives substantial simplification of the calculation. Also, since the number of spin wave modes are reduced, it facilitates our understanding of the type of correlation belonging to a certain spin wave mode. Also the formalism can provide a good starting point for higher order calculation in incommensurate structures as a function of $1/\sqrt{S}$. Recently it was shown how higher order terms in the spin wave expansion can lead to substantial magnon decay and finite lifetime in non-collinear magnets \cite{Zhitomirsky2012,Mourigal2013}. Additionally we generalize our method to arbitrary anisotropic exchange interactions including the Dzyaloshinskii-Moriya interaction.

  The algorithm of the proposed method is implemented in the open source Matlab toolbox called SpinW\cite{Toth2013}. The code can solve the linear spin wave problem both numerically and analytically.  
  
  The structure of the paper is the following. We introduce first the general magnetic Hamiltonian in Sec.\ \ref{sMagHam}, then we proceed step-by-step to produce the normal spin wave modes, dynamical structure factor and the sublattice magnetization. In Sec.\ \ref{sMag} the range of solvable magnetic ground state structures will be described. In order to solve incommensurate spin waves, the exchange interactions have to fulfill certain symmetries discussed in Sec.\ \ref{sSym}. Using the Holstein-Primakoff transformation, the magnetic Hamiltonian is transformed into a quadratic form of bosonic operators in Sec.\ \ref{sQuad} where also external magnetic field is introduced. The quadratic form is diagonalized using the Bogoliubov transformation in Sec.\ \ref{sDiag} with the less well known numerical method of Colpa [\onlinecite{Colpa1978}]. The diagonalized Hamiltonian contains the dispersion relations of the normal spin wave mode. In Sec.\ \ref{sCorr} the spin-spin correlation functions are extracted and the magnetization of each sublattice are calculated in Sec.\ \ref{sSubMag}. Sec.\ \ref{sCode} describes how the method can be converted into an algorithm. Finally, the general solution of the spin wave spectrum for magnetic C-site langasites is calculated in Sec.\ \ref{sLang} followed by a summary in Sec.\ \ref{sSum}. 

\section{Magnetic Hamiltonian}\label{sMagHam}

We would like to solve the most general magnetic Hamiltonian of interacting localized magnetic moments on a periodic lattice using LSWT. To accomplish this, a method is necessary that can deal with Hamiltonians where the quadratic spin exchange interactions are expressed with 3$\times$3 matrices. In this case the exchange energy of two spins will be a matrix product $\vect{S}_i^\intercal\matr{J}\vect{S}_j$, where $\vect{S}_i$ is a $3\times1$ column vector of the spin operators $\{S^x_i, S^y_i, S^z_i\}$ of site $i$ and $\matr{J}$ is the exchange matrix coupling the two sites. This matrix formalism includes the isotropic exchange (diagonal matrix), Dzyaloshinskii-Moriya exchange (antisymmetric matrix) and different anisotropic interactions (for example the Kitaev-exchange \cite{Kitaev2006}). The single ion anisotropy can be described in a similar manner using the $\vect{S}_i^\intercal\matr{A}\vect{S}_i$ expression. As an example easy-axis anisotropy along the $x$-axis is represented by a matrix, whose only non-zero element is the first diagonal with the negative easy axis energy. Similarly any local easy axis direction can be defined by the appropriate coordinate transformation of the anisotropy matrix. Including the external magnetic field and $g$-tensor, we propose to solve the following Hamiltonian:

\begin{align}
    \label{ham0}
  \matr{H} =& \sum_{\substack{mi\\nj}} \vect{S}_{mi}^\intercal\matr{J}_{mi,nj}\vect{S}_{nj} + \\ \nonumber
  &\sum_{mi} \vect{S}_{mi}^\intercal\matr{A}_{mi}\vect{S}_{mi} + \mu_B\vect{H}^\intercal \sum_{mi} \matr{g}_{i}\vect{S}_{mi}.
\end{align}
The indices $m$, $n$ are indexing the crystallographic unit cell (running from 1 to $L$), while $i$ and $j$ label the magnetic atoms inside the unit cell (running from 1 to $N$), $\vect{H}$ is the external magnetic field column vector, $\mu_B$ is the Bohr magneton. This Hamiltonian can describe the magnetic properties of many Mott insulators.

\section{General idea of the solution}\label{sGen}

S.\ Petit [\onlinecite{Petit2011}] calculated the general solution of Eq.\ \ref{ham0} for commensurate magnetic ground state by introducing a local coordinate transformation for every magnetic atom in the unit cell. This effectively transforms the ground state into ferromagnetic order where the spin wave spectrum is readily calculable. To solve models with an incommensurate ground state, we introduce a preceding coordinate transformation, the rotating frame \cite{Kaplan1961,Chernyshev2009}. It uniformly rotates the magnetic moments in every unit cell by an angle that depends on the magnetic ordering wave vector and the position of the cell transforming the magnetic order into a commensurate one. If the incommensurate magnetic structure can be transformed to a ferromagnetic one with these two subsequent rotations, then the spin wave spectrum will contain a finite number of well defined modes and can be solved by our method. Among the simplest examples of incommensurate magnetic structures are the 120$^\circ$ order of the isotropic triangular lattice antiferromagnet or the helical structure of the $J_1$-$J_2$ antiferromagnetic chain model. If the proposed two rotations cannot be constructed, then the spin wave Hamiltonian will contain umklapp terms, that couple magnons with different momentum and the spin wave spectrum will contain a continuum of states. An example of such a magnetic structure is two interacting counter rotating incommensurate spirals which form the ground state of $\beta$-CaCr$_2$O$_4$ \cite{Damay2010}. The existence of the above two rotations is intimately connected to the symmetry of the magnetic Hamiltonian that will be discussed in Sec.\ \ref{sSym}. The present method can be also applied for multi-Q magnetic structures, however in this case a $\vect{Q}=0$ magnetic supercell has to be constructed that incorporates the full magnetic structure (approximately for incommensurate structures).

\section{Magnetic ground state}\label{sMag}

In order to calculate the LSWT solution of the proposed Hamiltonian we need to determine its classical magnetic ground state. Acknowledging that this is often a challenging task, we assume that the solution is \textit{a priori} known. There is an extended literature on the determination of the classical magnetic ground state either using the Luttinger-Tisza method \cite{Luttinger1946a,Kaplan2007} or Monte-Carlo simulations \cite{Lacorre1987}. To parametrize the solvable magnetic structures, we use real vectors defining the classical direction of the spins denoted by $\vect{S}_{0j}$ in the first magnetic unit cell, while all other vectors $\vect{S}_{nj}$ are generated with a rotation of the vectors $\vect{S}_{0j}$ by the angle $\varphi_n$. The classical vector components will be substituted with the corresponding quantum mechanical spin operators in the Hamiltonian. The rotation angle depends on the magnetic ordering wave vector $\vect{Q}$ and the position of the magnetic cell $\vect{r}_n$:
\begin{align}
\varphi_n = \vect{Q}\cdot\vect{r}_n.
\end{align}
The classical spin direction of arbitrary site can be expressed as:
\begin{align}
  \vect{S}_{nj}  = \matr{R}_n\vect{S}_{0j},
  \label{Srot1}
\end{align}
where $\matr{R}_n$ is a rotation matrix, that depends only on the angle $\varphi_n$ about a global axis of rotation $\vect{n}$. On periodic crystals magnetic structures can be most conveniently expressed by Fourier coefficients:
\begin{align}
	\vect{S}_{nj} = \sum_\vect{k} \vect{F}_{\vect{k}j}\exp(-i\vect{k}\cdot\vect{r}_n).
\end{align}
Since the classical spin vectors are real vectors, the Fourier coefficients must fulfill the equality: 
\begin{align}
\vect{F}_{\vect{k}j} = \overline{\vect{F}}_{\vect{-k}j}.
\end{align}
It can be shown that the Fourier transform of Eq.\ \ref{Srot1} can have at most three Fourier components with wave vectors $\{0, \vect{Q}, -\vect{Q}\}$. We will call these magnetic structures a single-Q spin order. The ferromagnetic component $\vect{F}_{0j}$ has to be parallel to the global rotation axis $\vect{n}$, while the complex vectors $\vect{F}_{\pm\vect{Q}j}$ define the plane of the spin helix.

\begin{figure}[htbp]
	\centering
	\includegraphics[width = \linewidth]{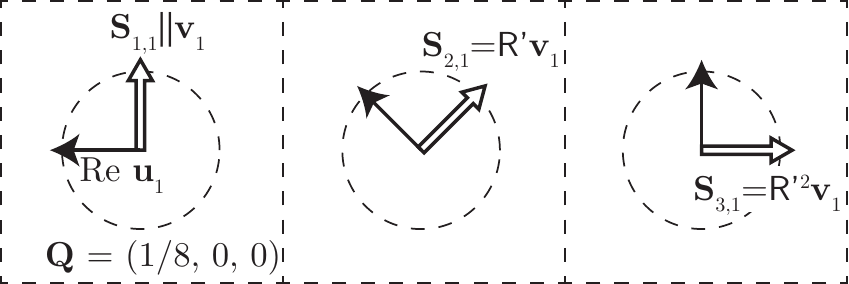}
	\caption{Rotating coordinate system of a single atom per unit cell magnetic helix, with ordering wave vector of $\vect{Q}=(1/8,0,0)$. The empty arrows denote the classical spin directions, the dashed squares show the crystallographic unit cells.}
	\label{fig:rot}
\end{figure}

To diagonalize the Hamiltonian we transform the classical spin vectors into a ferromagnetic order aligned parallel to the $z$-axis. The quantum mechanical spin operators will be transformed the same way, where fluctuations will be perpendicular to the local $z$-axis. First we change to the rotating frame. This defines a new set of operators $\vect{S}'_{nj}$:
\begin{align}
  \vect{S}_{nj}  = \matr{R}_n\vect{S}'_{nj}.
  \label{Srot}
\end{align}
The new vectors $\vect{S}'_{nj}$ will be independent of the $n$ index of the unit cell. A second coordinate transformation will rotate every magnetic moment within the unit cell to ferromagnetic order:
\begin{equation}
  \label{rotR}
  \vect{S}'_{nj} = \matr{R}'_j \vect{S}''_{nj}.
\end{equation}
The $\matr{R}'_j$  matrices describe local rotations which are independent of the position of the unit cell. The third column of $\matr{R}'_j$ is a unit vector pointing along the spin vector direction in the rotating frame, while the other two columns span an orthogonal coordinate system. The above matrix equation can be rewritten in the form of a sum:
\begin{equation}
 S'^\alpha_{nj} = \sum_\mu \matr{R}'^{\alpha \mu}_j S''^{\mu}_{nj},
\end{equation}
where $\alpha$ and $\mu$ runs over \(\{1,2,3\}\). Using the elements of the matrix $\matr{R}'_j$, two useful vectors can be defined:
\begin{eqnarray}
 u_j^\alpha &=& R'^{\alpha 1}_j + i R'^{\alpha 2}_j,\\
 v_j^\alpha &=& R'^{\alpha 3}_j,\nonumber
\end{eqnarray}
where $\vect{u}_j$ is complex vector and $\vect{v}_j$ is a unit vector parallel to the $j$th spin vector in the rotating frame, see Fig.\ \ref{fig:rot}.

\section{Symmetries of the Hamiltonian}\label{sSym}

In order to simplify the solution of Eq.\ \ref{ham0} the symmetries of the magnetic Hamiltonian need to be considered. Beside the lattice translation symmetry, the single-Q magnetic order requires that the magnetic Hamiltonian is invariant under the $\matr{R}_n$ rotations. These symmetries give constraints on the possible exchange matrices and anisotropies. From now on the $\matr{A}_{mi}$ anisotropy matrices will be merged into the $\matr{J}_{mi,mi}$ elements of the interaction matrices.

Due to the underlying periodic lattice, the exchange matrix has to be invariant under translations with arbitrary lattice vector:
\begin{equation}
\matr{J}_{mi,nj} = \matr{J}_{ij}(\vect{d}).
\end{equation}
The $\vect{d} = \vect{r}_n-\vect{r}_m$ is the lattice translation vector between the unit cells of the two interacting spins. 

The second symmetry is the invariance under exchange of the two interacting spins. In this case the $\matr{J}$ matrix has to be transposed, in order to reproduce the sign change of the antisymmetric exchange:
\begin{equation}
\matr{J}_{ij}(\vect{d}) = \matr{J}_{ji}^\intercal(-\vect{d}).
\end{equation}
The application of this symmetry ensures that the magnetic Hamiltonian will be Hermitian:
\begin{equation}
\matr{J}_{ij}(\vect{d}) = \frac{1}{2}\left(\matr{J}_{ij}(\vect{d}) + \matr{J}_{ji}^\intercal(-\vect{d})\right).
\label{Jsym}
\end{equation}
As a consequence all anisotropy matrices have to be symmetric. 

The third symmetry is the invariance under the rotation $\matr{R}_n$:
\begin{equation}
	\matr{J}_{ij}(\vect{d}) = \matr{R}_n^\intercal\matr{J}_{ij}(\vect{d})\matr{R}_n, \;n=1,2,3...
	\label{Jrot}
\end{equation}
Here we used the fact that the inverse of the rotation (orthogonal) matrix is its transpose. It will be useful to define the Fourier transform of the interaction matrices:
\begin{align}
  \matr{J}_{ij}(\vect{k}) = \sum_\vect{d} \matr{J}_{ij}(\vect{d}) e^{-i\vect{k\cdot d}}.
\end{align}
It is straightforward to determine the symmetries of $\matr{J}_{ij}(\vect{k})$:
\begin{eqnarray}
\matr{J}_{ij}(\vect{k}) &=& \overline{\matr{J}_{ij}(-\vect{k})}, \\ \nonumber
\matr{J}_{ij}(\vect{k}) &=& \overline{\matr{J}_{ji}^\intercal(\vect{k})}.
\end{eqnarray}

\section{Quadratic form}\label{sQuad}

In order to solve Eq. \ref{ham0} we apply linear spin wave theory. LSWT describes the dynamics of small fluctuations of the spins around their classical direction. As long as the expectation value of the spin operator is only weakly reduced from the classical value, the theory works well. This is typically true at low temperatures and large spins and LSWT is often a good approximation for systems with spin-3/2 and above while higher order corrections are certainly necessary for spin-1/2 systems. The expansion of the Hamiltonian as a function of $1/\sqrt{S}$ is achieved using the Holstein-Primakoff approximation \cite{Holstein1940}. The spin operators are expanded in terms of bosonic creation and annihilation operators on every magnetic site in the local coordinate system. By keeping only the lowest order of the boson operator we create a linear approximation of the complex spin dynamics:
\begin{eqnarray}
 S''^+_{nj} &=& \sqrt{2S_j} \;b_{nj}, \\\nonumber
 S''^-_{nj} &=& \sqrt{2S_j} \;b_{nj}^\dagger,\\\nonumber
 S''^z_{nj} &=& S_j - b_{nj}^\dagger b_{nj},
 \end{eqnarray}
 where $b^\dagger_{nj}$ and $b_{nj}$ decrease and increase the spin quantum number by one and fulfill the following bosonic commutation relations:
\begin{align}
[b_{mi},b^\dagger_{nj}] = \delta_{mn}\delta_{ij}.
\end{align} 
The real space components of the spins operators are the following:
 \begin{eqnarray}
 S''^1_{nj} &=& \frac{\sqrt{2S_j}}{2} ( b_{nj} + b_{nj}^\dagger),\\\nonumber
 S''^2_{nj} &=& \frac{\sqrt{2S_j}}{2i} ( b_{nj} - b_{nj}^\dagger),\\\nonumber
 S''^3_{nj} &=& S_i - b_{nj}^\dagger b_{nj}.
\end{eqnarray}
Using Eq.\ \ref{rotR} the spin operators in the rotating frame can be expressed with the bosonic operators as follows:
\begin{equation}
 \vect{S}'_{nj} = \sqrt{\frac{S_j}{2}} \left( \overline{\vect{u}}_j b_{nj} + \vect{u}_j b_{nj}^\dagger\right) + \vect{v}_j \left(
S_j - b_{nj}^\dagger b_{nj}\right).
\label{Sexpr}
\end{equation}
After substitution of Eq.\ \ref{Sexpr} and \ref{Srot} into the Hamiltonian one gets (without the magnetic field):
\begin{widetext}
  \begin{equation}
 \matr{H} = \displaystyle\sum_{\substack{mi\\nj}} \left\{ \sqrt{\frac{S_i}{2}}( \overline{\vect{u}}^\intercal_i b_{mi} + \vect{u}_i^\intercal b_{mi}^\dagger) +
\vect{v}_i ^\intercal(S_i-b_{mi}^\dagger b_{mi})\right\} 
\matr{R}_m^\intercal\matr{J}_{mi,nj} \matr{R}_n
\left\{ \sqrt{\frac{S_j}{2}}( \overline{\vect{u}}_j b_{nj} + \vect{u}_j b_{nj}^\dagger ) + \vect{v}_j (S_j-b_{nj}^\dagger b_{nj} )\right\}.
\label{ham}
  \end{equation}
  \end{widetext}
After expanding the right side in terms of increasing boson operator number, the first term is constant that gives the classical ground state energy. The expectation value of the one operator term vanishes and the two operator term gives the spin wave dispersion. In the linear approximation the higher order terms are neglected. The $\matr{R}_m^\intercal\matr{J}_{mi,nj} \matr{R}_n$ term describes a rotation of the interaction matrix depending on the unit cell indices of the interacting magnetic moments. Using the symmetry in Eq.\ \ref{Jrot}, new $\matr{J}'$ matrices can be defined:
\begin{equation}
  \matr{J}'_{mi,nj} = \matr{R}_m^\intercal\matr{J}_{mi,nj} \matr{R}_n = \matr{J}_{mi,nj} \matr{R}_{n-m}.
\end{equation}
It can be shown, that $\matr{J}'$ has the same symmetries as $\matr{J}$. In order to diagonalize the two operator expression the bosonic operators have to be Fourier transformed. The inverse transformation is: 
\begin{eqnarray}
 b_{mi} &=& \frac{1}{\sqrt{L}} \sum_{\vect{k} \in \textrm{B.Z.}} b_i(\vect{k}) e^{i\vect{k}\vect{r}_m},
\end{eqnarray}
where the summation runs over the first Brillouin zone. The two operator terms can be expressed in matrix form:
\begin{equation}
 H = \sum_{\vect{k}\in \textrm{B.Z.}}\vect{x}^\dagger(\vect{k}) \matr{h}(\vect{k}) \vect{x}(\vect{k}),
\label{hq}
\end{equation}
where $\vect{x}$ is the column vector of the bosonic operators:
\begin{equation}
\vect{x}(\vect{k}) = \left[ b_{1}(\vect{k}),\ldots,b_{N}(\vect{k}),b_{1}^\dagger(-\vect{k}),\ldots,b_{N}^\dagger(-\vect{k})\right]^\intercal.
\end{equation}

The Hermitian matrix $\matr{h}(\vect{k})$ consists the following sub-matrices:
\begin{equation}
  \label{grand}
 \matr{h}(\vect{k}) = \left[ \begin{array}{cc} \matr{A}(\vect{k}) - \matr{C} & \matr{B}(\vect{k}) \\
      \matr{B}^\dagger(\vect{k}) & \overline{\matr{A}}(-\vect{k}) - \matr{C}\end{array} \right],
\end{equation}
that contain the following $(i,j)$ elements:
\begin{eqnarray}
\label{abc}
A({\vect{k}})^{i,j} &=& \frac{\sqrt{S_iS_j}}{2} \vect{u}^\intercal_i\matr{J}'_{ij} (-\vect{k}) \vect{\overline{u}}_j, \\ \nonumber
B(\vect{k})^{i,j}   &=& \frac{\sqrt{S_iS_j}}{2} \vect{u}^\intercal_i\matr{J}'_{ij} (-\vect{k}) \vect{u}_j,\\ \nonumber
C(\vect{k})^{i,j}   &=& \delta_{ij}  \sum_l  S_l\vect{v}_i^\intercal\matr{J}'_{il}(0)\vect{v}_l.
\end{eqnarray}
It can be shown that $\matr{A}(\vect{k})$ is Hermitian and $\matr{C}$ is real.

To introduce the effect of external magnetic field, we also express the Zeeman term using the bosonic operators. After following the same steps as above, the external field energy in the rotating frame is the following:
\begin{align}
  \matr{H}^Z = -\mu_B\vect{H}^\intercal \sum_{\vect{k}j}\matr{g}_j\vect{v}_jb^\dagger_j(\vect{k})b_j(\vect{k}).
\end{align}
To avoid umklapp terms in the Hamiltonian, the $\vect{H}^{eff}_i = \matr{g}_i^\intercal\vect{H}$ effective field vector has to be invariant under the $\matr{R}_n$ rotations. This constrains the effective magnetic field to be parallel to the $\vect{n}$ global rotation axis. This Zeeman term has to be added to the $\matr{A}(\vect{k})$ matrix with the following elements:
\begin{align}
A^Z(\vect{k})^{i,j} = -\frac{1}{2} \mu_B \delta_{ij} \vect{H}^\intercal\matr{g}_i\vect{v}_i.
\end{align}

\section{Diagonalization of the quadratic form}\label{sDiag}

In order to determine the spectrum of the quadratic Hamiltonian we need to diagonalize the $\matr{h}(\vect{k})$ square matrices. Although $\matr{h}(\vect{k})$ is Hermitian, a simple unitary transformation is not sufficient, since the transformed $b'_i$ operators have to fulfill the bosonic commutation relations as well. This can be only achieved, if $\matr{h}(\vect{k})$ is positive definite \cite{Colpa1978}, as follows from the fact that the spectrum of the $\matr{H}$ Hamiltonian has a lower bound. In this case it can be shown, that the diagonalized Hamiltonian has only positive real numbers in the diagonal, that are doubly degenerate. An elegant solution to the diagonalization of a bosonic Hamiltonian is proposed by J. H. P. Colpa [\onlinecite{Colpa1978}], we describe his method briefly in the following. 

We express the commutation relations of the $b_i$ operators in a matrix form:
\begin{equation}
 \left[\vect{x},\vect{x}^\dagger\right] = \vect{x}(\vect{x}^*)^\intercal - (\vect{x}^* \vect{x}^\intercal)^\intercal = \matr{g},
\end{equation}
where $\vect{x}^*$ is the column matrix of the Hermitian adjoint operators ($\matr{g}$ is not to be confused with the $\matr{g}_i$ atomic g-tensors). These commutation relations have to be fulfilled by the new bosonic operators that create the normal spin wave modes. Using the previously defined value of $\vect{x}$, the value of the commutator matrix is the following:
\begin{equation}
 \matr{g} = \left[ \begin{array}{cc} \matr{1} & 0 \\ 
      0 & -\matr{1}\end{array} \right],
\end{equation}
where $\matr{1}$ is the identity matrix, with dimensions of $N\times N$. As the first step of the solution, the Cholesky decomposition has be applied on $\matr{h}(\vect{k})$ to find the $\matr{K}$ complex matrix that fulfills the following equation (implicitly assuming the $\vect{k}$ dependence):
\begin{align}
  \matr{h}(\vect{k}) = \matr{K}^\dagger\matr{K}.
\end{align}
Afterwards the eigenvalue problem of the Hermitian $\matr{KgK}^\dagger$ matrix has to be solved. The resulting $\vect{f}_i$ eigenvectors are arranged into the matrix $\matr{U}$ as column vectors in such a way that the first $N$ diagonal elements of the diagonalized $\matr{L} = \matr{U}^\dagger \matr{KgK}^\dagger \matr{U}$ matrix are positive and the last $N$ elements are negative. The diagonal matrix is then given by:
\begin{align}
  \matr{E} = \matr{gL},
\end{align}
where the first $N$ diagonal elements $E_i(\vect{k}) := E_{ii}$ are the energies of the normal spin wave modes that are the function of the wave vector $\vect{k}$. The second $N$ eigenvalues are equal to the first $N$ multiplied by minus one. Each boson mode is a linear combination of the $b_j'$ and $b'^\dagger_j$ normal modes:
\begin{equation}
 x_i=\sum_{j} T_{ij}x_j',
\end{equation}
where the transformation matrix $\matr{T}$ can be calculated as:
\begin{align}
  \matr{T} = \matr{K}^{-1}\matr{UE}^{1/2}.
\end{align}
In case the spectrum of the Hamiltonian $\matr{H}$ contains zero energy modes (e.g. Goldstone modes), the $\matr{h}(\vect{k})$ matrix will be positive semidefinite for certain $\vect{k}$ values. This can be cured by adding a small positive $\epsilon$ value to the diagonal of $\matr{h}(\vect{k})$. It introduces only a negligible gap in the spectrum, but makes the $\matr{h}(\vect{k})$ matrix positive definite and the decomposition problem solvable.

\section{Dynamical correlation functions}\label{sCorr}

Beside the spin wave dispersion another measurable quantity is the spin-spin correlation function. This can be directly measured by inelastic neutron scattering as a function of momentum and energy transfer \cite{Marshall1971}. The dynamical correlation function can be expressed as a $3\times3$ matrix as a function of momentum and energy:
\begin{widetext}
  \begin{equation}
 \matr{S}(\vect{k},\omega) = \frac{1}{2\pi N} \sum_{\substack{mi\\nj}}
e^{i\vect{k}(\vect{r}_{mi}-\vect{r}_{nj})}\int_{-\infty}^{\infty} \textrm{d}\tau\; e^{-i\omega \tau} \langle \vect{S}_{mi} \vect{S}_{nj}^\intercal
(\tau)\rangle,
\label{corf}
  \end{equation}
\end{widetext}
where $\vect{r}_{mi}$ is the position vector of the magnetic atoms that can be expressed in terms of the relative position vector $\vect{t}_i$ of atom $i$ and the position vector of the $m$th unit cell $\vect{r}_m$:
\begin{align}
\vect{r}_{mi} = \vect{r}_m+\vect{t}_i.
\end{align}
Using Eq.\ \ref{Srot} the real space-time spin-spin correlation function in the laboratory frame can be expressed as:
\begin{align}
   \label{SSrot}
   \langle \vect{S}_{mi}\vect{S}_{nj}^\intercal (\tau)\rangle &= \langle \matr{R}_m\vect{S}_{mi}'\vect{S}_{nj}'^\intercal(\tau)\matr{R}_n^\intercal\rangle  \\ &=\langle\vect{S}_{mi}'\vect{S}_{nj}'^\intercal (\tau)\rangle\matr{R}_{n-m}^\intercal,\nonumber
\end{align}
using the fact that the correlation function is invariant under a shift of the origin by any lattice vector except when $2\vect{Q}=\tau$. Since the calculated $\langle\vect{S}_{mi}'\vect{S}_{nj}'^\intercal (\tau)\rangle$ expression is not necessarily invariant under the rotation $\matr{R}_n$ due to the arbitrary choice of the zeroth cell. The symmetrization (denoted by $\langle...\rangle_\matr{R}$) can be achieved by the following integral:
\begin{align}
\langle\vect{S}_{mi}'\vect{S}_{nj}'^\intercal (\tau)\rangle_{\matr{R}} = \int_{\varphi=0}^{2\pi} \langle\vect{S}_{mi}'\vect{S}_{nj}'^\intercal (\tau)\rangle\matr{R}(\varphi).
\end{align}

To perform the Fourier transform on this expression, the $\matr{R}_n$ matrices have to be split into different periodic components as a function of the lattice translation vector $\vect{r}_n$. This can be achieved using Rodrigues' formula:
\begin{align}
  \matr{R}(\vect{Q}\cdot \vect{r}_n) &= e^{ i\vect{Q}\cdot \vect{r}_n}\matr{R}_1 + e^{-i\vect{Q}\cdot \vect{r}_n}\overline{\matr{R}}_1 + \matr{R}_2,\\\nonumber
  \matr{R}_1 &= 1/2 \;\left(\matr{1} - i\left[\vect{n}\right]_\times-\vect{n}\vect{n}^\intercal\right),\\\nonumber
  \matr{R}_2 &= \vect{n}\vect{n}^\intercal,\\\nonumber
  \left[\vect{n}\right]_\times &= \left[\begin{array}{ccc}0 & -n_z&n_y\\n_z& 0 & -n_x\\-n_y & n_x & 0\end{array}\right].
\end{align}
After substituting it into Eq.\ \ref{corf}, one gets:
\begin{align}
  \matr{S}(\vect{k},\omega) =& \matr{S}'(\vect{k},\omega)\matr{R}_2 + \matr{S}'(\vect{k}+\vect{Q},\omega)\matr{R}_1 +\\ &+\matr{S}'(\vect{k}-\vect{Q},\omega)\overline{\matr{R}}_1,\nonumber
\end{align}
where $\matr{S}'(\vect{k},\omega)$ is the correlation function in the rotating frame calculated from $\vect{S}'_{nj}$ operators:
\begin{align}
    \matr{S}'(\vect{k},\omega) =\frac{1}{2\pi}\int_{-\infty}^\infty\textrm{d}\tau\; e^{-i\omega \tau} \matr{S}'(\vect{k},\tau).
\end{align}
In this form it is clear that the correlation function of incommensurate spin structures has a magnon dispersion at $\omega(\vect{k}\pm\vect{Q})$ in addition to that at $\omega(\vect{k})$. In case the magnetic atoms are on a Bravais lattice, the $\matr{S}'(\vect{k},\omega)$ correlation describe rigid rotation of the spins in the ordering plane, this mode is called phason, while the $\matr{S}'(\vect{k}\pm\vect{Q},\omega)$ correlations describe the canting of the spins away from the ordering plane \cite{Coldea2003}.

The $\matr{S}'(\vect{k},\omega)$ correlation functions can be calculated, using the definition of Eq.\ \ref{Sexpr} and keeping only the two operator terms. Four operator terms appearing in the correlation function are responsible for longitudinal fluctuation of the spins. This leads to a continuum of two magnon scattering, that is disregarded here but also calculable using our framework. As a first step, the spatial Fourier transform is calculated keeping only the two operator terms:
\begin{widetext}
\begin{align}
S'^{\alpha\beta}(\vect{k},\tau) =& \frac{1}{N}\sum_{ij}\Bigg\{e^{i\vect{k}(\vect{t}_i-\vect{t}_j)}\frac{\sqrt{S_i S_j}}{2} \langle\left[b_i^\dagger(\vect{k}),b_i(-\vect{k})\right]  \left[
\begin{array}{cc} u_i^\alpha \overline{u}_j^\beta & u_i^\alpha u_j^\beta\\
 \overline{u}_i^\alpha \overline{u}_j^\beta & \overline{u}_i^\alpha u_j^\beta\end{array} \right] 
      \left[\begin{array}{c} b_j(\vect{k},\tau)\\b_j^\dagger(-\vect{k},\tau)\end{array} \right]\rangle-\\
&-\delta(\vect{k}-\boldsymbol{\kappa}) \nu_i^\alpha \nu_j^\beta \sum_\vect{k'}\langle S_i b_j^\dagger(\vect{k'})b_j(\vect{k'})+S_j b_i^\dagger(\vect{k'})b_i(\vect{k'})\rangle\Bigg\}.\nonumber
\end{align}
\end{widetext}
The first term describes a time dependent scattering process, while the second term describes the reduction of the static ordered moment due to magnon population. The $\delta(\vect{k}-\boldsymbol{\kappa})$ expression is non-zero at the $\boldsymbol{\kappa}$ reciprocal lattice vectors in the rotating coordinate system, that are identical to the magnetic Bragg peak positions in a lab coordinate system. The dynamical part of the correlation function in matrix form is:
\begin{align}
  S'^{\alpha\beta}(\vect{k},\tau) = \frac{1}{2N} \langle\vect{x}^\dagger(\vect{k})\left[\begin{array}{cc} \matr{Y}^{\alpha\beta} & \matr{Z}^{\alpha\beta} \\ \matr{V}^{\alpha\beta} & \matr{W}^{\alpha\beta}\end{array}\right]\vect{x}(\vect{k},\tau)\rangle.
\end{align}
This is a sum of the expectation values of boson operator pairs with the following coefficients, the $(i,j)$ elements of the four sub matrices with dimensions of $N\times N$:
\begin{align}
  \left[Y^{\alpha\beta}\right]^{i,j} &= \sqrt{S_iS_j}\;e^{i\vect{k}(\vect{t}_i-\vect{t}_j)}u_i^\alpha \overline{u}_j^\beta,\\\nonumber
  \left[Z^{\alpha\beta}\right]^{i,j} &= \sqrt{S_iS_j}\;e^{i\vect{k}(\vect{t}_i-\vect{t}_j)}u_i^\alpha u_j^\beta,\\\nonumber
  \left[V^{\alpha\beta}\right]^{i,j} &= \sqrt{S_iS_j}\;e^{i\vect{k}(\vect{t}_i-\vect{t}_j)}\overline{u}_i^\alpha \overline{u}_j^\beta,\\\nonumber
  \left[W^{\alpha\beta}\right]^{i,j} &= \sqrt{S_iS_j}\;e^{i\vect{k}(\vect{t}_i-\vect{t}_j)}\overline{u}_i^\alpha u_j^\beta.
\end{align}
The expectation value of the new bosonic operators must also be determined in order to evaluate the term within the brackets $\langle \vect{x}^\dagger ... \vect{x}\rangle$:
\begin{align}
   \label{expect}
 \langle b_i^{\prime\dagger}(\vect{k}) b'_j(\vect{k},\tau)\rangle =&\; \delta_{ij}n(\omega_i(\vect{k}))e^{-i\omega_i (\vect{k})\tau},\\
 \langle b'_i(\vect{k}) b_j^{\prime\dagger}(\vect{k},\tau)\rangle =&\; \delta_{ij}(1+n(\omega_i(\vect{k})))e^{i\omega_i(\vect{k})\tau},\nonumber
\end{align}
where $\omega_i = E_i/\hbar$ and $n(\omega_i)$ is the Bose factor at temperature $T$:
\begin{align}
  n(\omega_i) = \frac{1}{e^{\hbar\omega_i/k_BT}-1}.
\end{align}
After substituting $\vect{x}(\vect{k})$ with the normal boson operators and performing the temporal ($\tau$) Fourier transformation of the dynamical correlation functions, one gets the final expression:
\begin{widetext}
\begin{align}
  \label{corf2}
  S'^{\alpha\beta}(\vect{k},\omega) = \frac{1}{2N}\sum_{i=1}^{2N} \left[ \matr{T}^\dagger \left[\begin{array}{cc} \matr{Y}^{\alpha\beta} & \matr{Z}^{\alpha\beta} \\ \matr{V}^{\alpha\beta} & \matr{W}^{\alpha\beta}\end{array}\right]\matr{T}\right]_{ii} \delta(\omega-g_{ii}\omega_i) \left(n(\omega)+\frac{1}{2}(1-g_{ii})\right).
\end{align}
\end{widetext}
In this equation the $g_{ii}$ diagonal elements of the commutation matrix $\matr{g}$ are used to produce the right magnon populations for the diagonal terms, it is also assumed that $g_{ii}\omega_i$ is sorted in decreasing order with the mode index $i$. Since all $\omega_i$ eigenvalues are positive, there will be $N$ positive and $N$ negative energies in the correlation function expression. To get an overview, we substitute Eq.\ \ref{corf2} into the neutron scattering cross section formula. Then it is clear, that the first $N$ $\langle b_i^\dagger b_i\rangle$ expectation values describe the probability of a neutron absorbing one magnon, while the $\langle b_ib_i^\dagger\rangle$ terms describe the magnon creation process. Since $\matr{S}'(\vect{k},\omega)$ has $N$ spin wave modes, a general incommensurate spin structure will have $3N$ measurable spin wave modes. 

\section{Sublattice magnetization}\label{sSubMag}

Linear spin wave theory also gives the leading correction to the size of the sublattice magnetization. This is reduced from the single ion moment value due to zero point quantum fluctuations and thermally excited spin waves at $T>0$. The sublattice magnetization reduction is independent of the moment size. The absolute value of the reduced moment:
\begin{align}
  |\langle\vect{S}_j\rangle| = \frac{1}{L}\sum_n\left|\langle\vect{S}_{nj}\rangle\right| = S_j - \frac{1}{L}\sum_n\langle b_{nj}^+b_{nj}\rangle.
\end{align}
The above summation can be accomplished using the Fourier transformed bosonic operator (the $\vect{k}$ sum runs over the first Brillouin zone):
\begin{align}
  \delta S_j &= -\frac{1}{L}\sum_{\vect{k} \in \textrm{B.Z.}} \langle b_j^+(\vect{k})b_j(\vect{k})\rangle \\
             &= -\frac{1}{L}\sum_{\vect{k} \in \textrm{B.Z.}}\langle\vect{x}(\vect{k})\vect{x}(\vect{k})^\dagger)\rangle_{j+N,j+N},\nonumber
\end{align}
where $\langle\rangle_{j+N,j+N}$ denotes the diagonal elements of the dyadic matrix containing the expectation values. Using the $\matr{T}$ transformation matrix and the expectation values of the normal boson operator pairs (Eq.\ \ref{expect}) the result is the following matrix equation:
\begin{align}
  \delta S_j &= -\frac{1}{L} \sum_{\vect{k} \in \textrm{B.Z.}} \left\{\matr{T}(\vect{k})\matr{D}(\vect{k})\matr{T}(\vect{k})^\dagger\right\}_{j+N,j+N},
\end{align}
where $\matr{D}(\vect{k})$ is a diagonal matrix that contains the $\langle\vect{x}(\vect{k})'\vect{x}(\vect{k})^{\prime\dagger}\rangle$ expectation value of the normal bosonic operators.

\section{Algorithm}\label{sCode}

In this section, we show how the above described general solution is implemented in SpinW\cite{Toth2013}. The input parameters of the calculation are the couplings and the magnetic structure. The couplings are stored in a list together with the anisotropy matrices. Every coupling is defined by several parameters. For the $l$th coupling $\vect{d}_l$ gives the distance vector between the origin of the unit cells of the interacting atoms, $i_l$ and $j_l$ are the indices of the interacting atoms and $\matr{J}_l$ is the $3\times3$ matrix of the interaction. For single ion anisotropy $\vect{d}_l=0$ and $i_l=j_l$. Thus the input list is the following $\{\vect{d}, i, j, \matr{J}\}_l$. To unambiguously define the magnetic structure, we need the classical spin direction of the $N$ magnetic atoms as $\vect{S}_i$ vectors, the $\vect{Q}$ ordering wave vector and the $\vect{n}$ normal vector. The first step of the calculation is to define a local Descartes coordinate system with axes $\{\vect{e}^1, \vect{e}^2, \vect{e}^3\}_i$ for each $\vect{S}_i$ classical spin. As the $\vect{e}^3_i$ vector is parallel to the spin direction, the $\vect{v}_i$ and complex $\vect{u}_i$ vectors are defined as:
\begin{align}
  \vect{u}_i &= \vect{e}^1_i + i\vect{e}^2_i,\\
  \vect{v}_i &= \vect{e}^3.\nonumber
\end{align}
For incommensurate structures one has to calculate the $\matr{J}'_l$ matrices by multiplying $\matr{J}_l$ matrices on the right side with a rotation matrix that rotates around $\vect{n}$ by the angle $\varphi=2\pi\vect{Q}\cdot\vect{d}$. To calculate the spin wave spectrum at $\vect{k}$ reciprocal space position, we need to calculate the $\matr{A}$, $\matr{B}$, $\matr{C}$ matrices. To calculate these, we run a summation over the $l$ index of the list of couplings. Each $l$ value is associated with an $(i,j)$ index representing the indices of the interacting atoms in the unit cell. Thus the $(i,j)$ element of $\matr{A}$, $\matr{B}$, $\matr{C}$ get an additional term according to Eq.\ \ref{abc} where $\matr{J}'(\vect{k})_l$ is simply determined by the following equation:
\begin{align}
\matr{J}'(\vect{k})_l=\matr{J}'_l\exp(2\pi\vect{k}\cdot\vect{d}_l).
\end{align}
Finally the vector of spin wave energies at $\vect{k}$ can be calculated from the $\matr{h}(\vect{k})$ matrix according to Section \ref{sDiag}. The algorithm is available as an open source code\cite{Toth2013}.

\section{Linear spin wave theory of B\lowercase{a}$_{3}$N\lowercase{b}F\lowercase{e}$_{3}$S\lowercase{i}$_{2}$O$_{14}$}\label{sLang}

\begin{figure}[htbp]
	\centering
	\includegraphics[width = \linewidth]{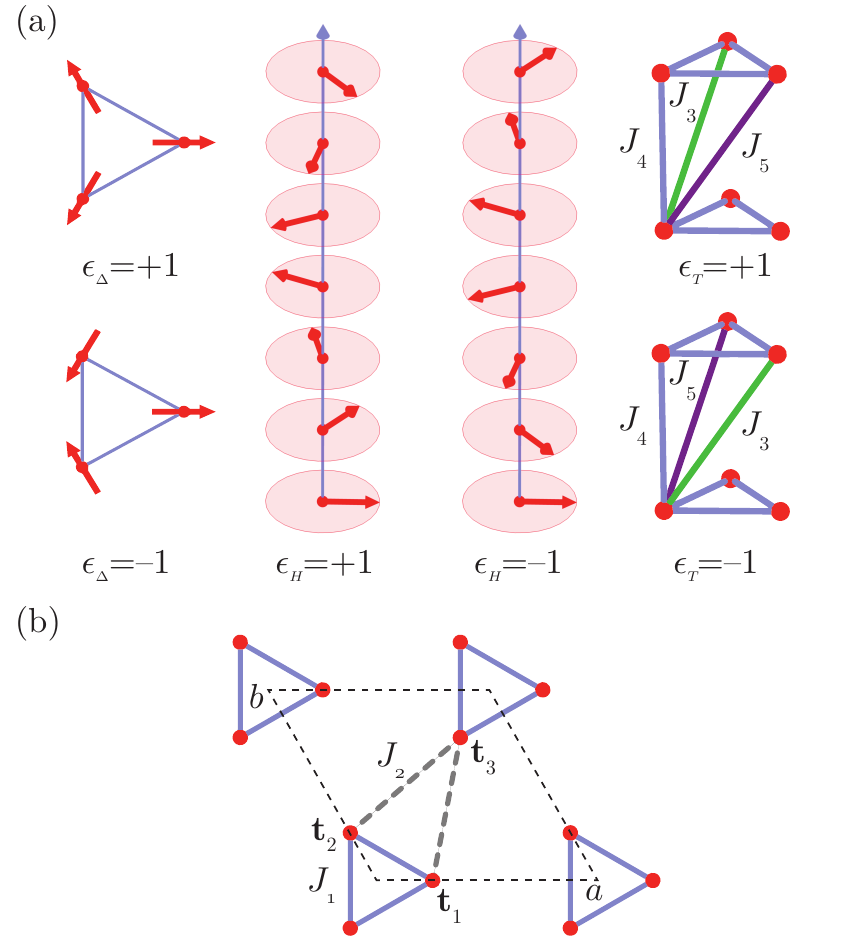}
	\caption{(a) The three different chiral property defined for the magnetic C-site langasites, (b) exchange couplings and positions of the magnetic ions in the $ab$-plane, where $J_1$ interactions define the triangular units.}
	\label{fig:chiral}
\end{figure}

As an example, we show a general model of the spin wave spectrum for magnetic compounds in the langasite family with incommensurate magnetic order. The langasite family with the prototype compound La$_3$Ga$_5$SiO$_{14}$ has non-centrosymmetric space group $P321$ and members of this family are being extensively studied due to their interesting piezoelectric and nonlinear optical properties \cite{Iwataki2001,Fritze2001}. Their general chemical formula is A$_3$BC$_3$D$_2$O$_{14}$ which contains four different cation sites, making it possible to accommodate several different magnetic ions. If the magnetic atoms occupy the A site, they build up stacked kagome layers, such as R$_3$Ga$_5$SiO$_{14}$ with R$=$Nd, Pr \cite{Bordet2006}. If the C site is occupied, the geometry of the interactions is more complex. The magnetic sites build up triangles which are themselves organized into a triangular lattice in the $ab$ plane and stacked along the $c$-axis, see Fig.\ \ref{fig:chiral}(b). The most studied C-site magnetic compounds contain Fe$^{3+}$ ions that have spin only magnetic moment $S=5/2$ \cite{Marty2008,Marty2010}. Marti et al.\ determined the magnetic structure of four different compositions using neutron diffraction combining A=Ba, Sr with B=Nb, Sb and D=Si. All four compositions have similar magnetic structures with incommensurate magnetic ordering wave vector $\vect{Q} = (0,0,\tau)$, see Tab.\ \ref{tab:tau}. The moments are oriented in the $ab$-plane and the angle between neighbors on the triangular units are 120$^\circ$. The magnetic ground state and excitation spectrum of Ba$_3$NbFe$_3$Si$_2$O$_{14}$ was studied in great detail by polarized and unpolarized inelastic neutron scattering \cite{Marty2008,Stock2011,Loire2011}. The modeling of the spectrum were done using a random phase approximation based on interacting trimers\cite{Jensen2011} and linear spin wave theory. However the published linear spin wave theory calculation assumed an ordering wave vector of $\tau=1/7$ and modeled the spin wave spectrum on a magnetic supercell with 7 unit cells along the $c$-axis. This cannot be generalized for the different $\tau$ values of the other compounds. Here we will present how our general linear spin wave theory can be used to model the spectrum of the other members of the langasite family. 

\begin{ruledtabular}
	\begin{table}[!htb]
		\centering
		\caption{Magnetic ordering wave vector of selected langasite compounds \cite{Marty2010}.}
		\label{tab:tau}
		\begin{tabular}{cc}
		Langasite & Ordering wave vector \\
		\hline
		Ba$_3$NbFe$_3$Si$_2$O$_{14}$ & $(0,0,0.1429(2))$\\
		Sr$_3$NbFe$_3$Si$_2$O$_{14}$ & $(0,0,0.1398(3))$\\
		Ba$_3$SbFe$_3$Si$_2$O$_{14}$ & $(0,0,0.1957(1))$\\
		Sr$_3$SbFe$_3$Si$_2$O$_{14}$ & $(0,0,0.1769(2))$\\
   \end{tabular}
\end{table}
\end{ruledtabular}

Due to the missing inversion symmetry, the crystal structure of Ba$_3$NbFe$_3$Si$_2$O$_{14}$ is chiral. This chirality is denoted by $\epsilon_T$ that can have values of $\pm 1$. This feature also appears in the geometry of the exchange pathways where $J_3$ and $J_5$ can have different values, see Fig.\ \ref{fig:chiral}(a). If $\epsilon_T=1$ the $J_3$ coupled chains have a right handed rotation along the $c$-axis, while the $J_5$ chains are left handed, for $\epsilon_T=-1$ the two are exchanged. The magnetic ground state also has chiral properties, the triangular units can have $\epsilon_\Delta=\pm 1$ chirality and the helical structure along the $c$-axis can have two sense of rotations $\epsilon_H=\pm 1$ for right handed/left handed spiral. Assuming that $J_5>J_3, J_4$, these three chiralities are related \cite{Marty2008}:
\begin{align}
\epsilon_T = \epsilon_\Delta \epsilon_H.
\end{align}
Thus for a certain $\epsilon_T$ crystal chirality additional antisymmetric exchange interactions are necessary to determine the sign of $\epsilon_\Delta$ and $\epsilon_H$.

\begin{ruledtabular}
	\begin{table}[!htb]
		\centering
		\caption{Positions and local coordinate system for the Fe$^{3+}$ magnetic ions, $\vect{t}_i$ gives the idealized atomic positions in lattice units and the $\vect{v}_i$ and $\vect{u}_i$ vectors define the magnetic structure using the $\ede$ chirality of the triangle units.}
		\label{tab:fepos}
		\begin{tabular}{p{0.2cm}llll}
		$i$ & $\vect{t}_i$   & $\vect{v}_i=\vect{S}_{0i}/|\vect{S}_{0i}|$  & $\vect{u}_i$ \\
		\hline
		1   & $(1/4,0,1/2)$   & $(1,0,0)$                 & $(0,1,i)$                     \\
		2   & $(0,1/4,1/2)$   & $(1/2,\sqrt{3}/2\ede,0)$  & $(\sqrt{3}/2,1/2\ede,-i\ede)$ \\
		3   & $(3/4,3/4,1/2)$ & $(1/2,-\sqrt{3}/2\ede,0)$ & $(\sqrt{3}/2,-1/2\ede,i\ede)$ \\
   \end{tabular}
\end{table}
\end{ruledtabular}

To model the spin wave spectrum we omit the weak but necessary antisymmetric exchange interaction, that can be included in a straightforward manner. The magnetic structure is described by the ordering wave vector $\vect{Q}=(0,0,\ehe\tau)$, the normal vector $\vect{n}=(0,0,1)$ and magnetic moment directions shown in Tab.\ \ref{tab:fepos}. The magnetic moment directions define the $\vect{u}_i$ and $\vect{v}_i$ vectors, where the complex $\vect{u}_i$ depends on the choice of coordinate system. The list of interactions is shown in Tab. \ref{tab:banbJlist}. Before generating the matrix of the Hamiltonian, one has to ensure that the interaction matrices fulfill all necessary symmetries defined in Sec.\ \ref{sSym} by applying Eq.\ \ref{Jsym}. To generate the $\matr{J}'$ transformed interaction matrices, we need to construct the $\matr{R}_{n-m}$ rotations. The $\matr{R}_{n-m}$ matrices introduce rotations around the $\vect{n}$ normal vector by the $\varphi_{n-m}=2\pi\vect{Q}\cdot\vect{r}_{n-m}$ angle that can have only two different values $\{0,\ehe\varphi_0\}$, where $\varphi_0=2\pi\tau$. The two rotation matrices are $\{\matr{1},\matr{R}_0\}$, where $\matr{R}_0$ rotates around the $c$-axis by $2\pi/7$ radians and has the following matrix elements:
\begin{align}
\matr{R}_0 = \left[\begin{array}{ccc} \cos(\varphi_0) & -\ehe\sin(\varphi_0) & 0 \\ \ehe\sin(\varphi_0) & \cos(\varphi_0) & 0 \\ 0 & 0 & 1 \end{array}\right].
\end{align}

After substituting the $\matr{J}_{ij}$ matrices into the formula of the quadratic form, we get to the following matrix form of the bosonic Hamiltonian:
\begin{align}
\matr{H} = \frac{S}{4}\left[\begin{array}{cccccc} A & B & E & H & \overline{F} & \overline{D}\\ \overline{B} & A & C & F & H & \overline{G} \\ \overline{E} & \overline{C} & A & D & G & H\\ H & \overline{F} & \overline{D} & A & B & E \\ F & H & \overline{G} & \overline{B} & A & C \\ D & G & H & \overline{E} & \overline{C} & A \end{array}\right].
\end{align}

\begin{figure}[htbp]
	\centering
	\includegraphics[width = \linewidth]{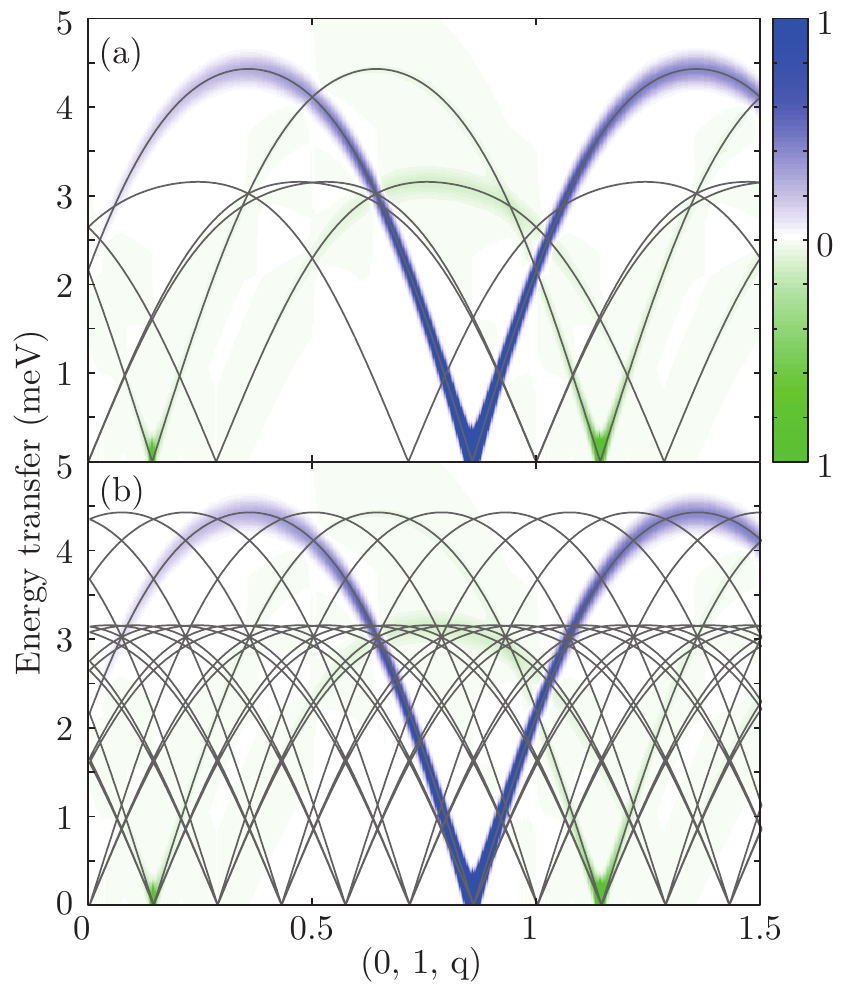}
	\caption{The calculated chiral correlation function $\operatorname{Im}(\matr{S}^{y'z'}-\matr{S}^{z'y'})$ of Ba$_{3}$NbFe$_{3}$Si$_{2}$O$_{14}$ assuming the same exchange parameters as Loire et al.\ \cite{Loire2011}. (a) the correlation function is calculated with the incommensurate model, and only $\omega(\vect{k}\pm\vect{Q})$ dispersions are plotted as lines, (b) the correlation function is calculated using the commensurate 7 unit cell model \cite{Loire2011}. Gray lines denote the spin wave dispersion, while the green to blue colors denote the intensity convoluted in energy with a 0.25 meV FWHM Gaussian.}
	\label{fig:C}
\end{figure}

The matrix elements as a function of $\vect{k}$ are the following:
\begin{widetext}
\begin{align}
A = \;&4J_1+8J_2+4J_3\gamma^- +4J_4\left(\cos(l)\left(\cos(\varphi_0)+1\right)-2\cos(\varphi_0)\right)+4J_5\gamma^+,\\ \nonumber
B = \;&-\ede\left(J_1+J_2(e^{ih}+e^{-ik})-J_3e^{-i\ete l}(\gamma^--2)-J_5e^{i\ete l}(\gamma^+-2)\right),\\ \nonumber
C = \;&-J_1e^{-i(h+k)}-J_2(1+e^{-ih})+J_3e^{-i(h+k+\ete l)}(\gamma^--2)+J_5e^{-i(h+k-\ete l)}(\gamma^+-2),\\ \nonumber
D = \;&-\ede G(k,h,-l),\\ \nonumber
E = \;&-\ede C(k,h,-l),\\ \nonumber
F = \;&-\ede\left(-3J_1-3J_2(e^{-ih}+e^{ik})-J_3e^{i\ete l}(\gamma^-+2)-J_5e^{-i\ete l}(\gamma^++2)\right), \\ \nonumber
G = \;&3J_1e^{i(h+k)}+3J_2(1+e^{ih})+J_3e^{i(h+k+\ete l)}(\gamma^-+2)+J_5e^{i(h+k-\ete l)}(\gamma^++2),\\ \nonumber
H = \;&4J_4\cos(l)\left(\cos(\varphi_0)-1\right).
\end{align}
\end{widetext}
where:
\begin{align}
\gamma^\pm = &\cos(\varphi_0)\pm\sqrt{3}\sin(\varphi_0).
\end{align}
The components of the momentum vector are denoted by $(h,k,l)=2\pi \vect{k}$ where $\vect{k}$ is in reciprocal lattice units. We used the relation between the magnetic and crystal chirality and accounted for $\ete=-1$ by exchanging $J_3$ and $J_5$. The $\gamma^\pm$ prefactor of $J_3$ and $J_5$ exchange interactions is related to the angles between the ordered moments on two triangular units on top of each other ($\sqrt{3}=2\sin(120^\circ)$). Unfortunately there is no short expression for the eigenvalues and eigenvectors of $\matr{gH}$, but they can be calculated using numerical methods. The eigenvalues of the $\matr{gH}$ matrix gives three positive spin wave energies, and together with the $\matr{S}'(\vect{k}\pm\vect{Q},\omega)$ terms in the spin-spin correlation function this model gives 9 possible spin wave modes. After calculating the spin-spin correlation functions according to Sec.\ \ref{sCorr}, we find that only six magnon modes are polarized in the $ab$ plane and the imaginary part (chiral part) of the correlation function has only $ab$-plane components:
\begin{align}
\matr{S}_{C}(\vect{k},\omega) = \operatorname{Im}\left(\matr{S}'(\vect{k}+\vect{Q},\omega)\matr{R}_1+\matr{S}'(\vect{k}-\vect{Q},\omega)\overline{\matr{R}}_1\right).
\end{align}
Whereas the $c$-axis polarized spin wave modes only have a contribution from $\operatorname{Re}\left(\matr{S}'(\vect{k},\omega)\matr{R}_2\right)$. The calculated correlation function $\operatorname{Im}(\matr{S}^{y'z'}-\matr{S}^{z'y'})$ is shown on Fig.\ \ref{fig:C}(a) and can be compared to the result of the Loire et al.\ on Fig.\ \ref{fig:C}(b). Here we used the $x'y'z'$ Cartesian coordinate system common for neutron scattering, where $x'$ is parallel to $\vect{k}$ and $y'$ is in the scattering plane. Both models give the same physically observable intensity, however our incommensurate model is more efficient since it gives only those magnon modes that have non-zero intensity. 

We expect that our spin wave model will be applicable to other incommensurate compounds in the langasite family with magnetic C-site such as the ones given in Tab.\ \ref{tab:tau} and we hope that our results stimulate further investigation of the dynamical magnetic properties of this interesting family.

\begin{ruledtabular}
	\begin{table}[!htb]
		\centering
		\caption{List of the exchange pathways for $\ete=1$ (for $\ete=-1$ the $J_3$ and $J_5$ interactions are exchanged), $\matr{J}$ matrices with the $i$ and $j$ indices of the interacting ions, $\vect{d}$ distance vector in lattice units and the $\matr{R}$ rotation matrices that define $\matr{J}'$.}		\label{tab:banbJlist}
		\begin{tabular}{lllll}
		Name & $i$ & $j$ & $\vect{d}$                 & $\matr{R}(\vect{d})$\\\hline
\multirow{3}{*}{$J_{1}$  }     &  1 &  2 & $( 0,  0,  0)$ & $1$            \\ 
                               &  2 &  3 & $(-1, -1,  0)$ & $1$            \\ 
                               &  3 &  1 & $( 1,  1,  0)$ & $1$            \\ \hline
\multirow{6}{*}{$J_{2}$  }     &  3 &  2 & $( 1,  0,  0)$ & $1$            \\ 
                               &  1 &  3 & $( 0,  0,  0)$ & $1$            \\ 
                               &  2 &  1 & $(-1,  0,  0)$ & $1$            \\ 
                               &  3 &  1 & $( 0,  1,  0)$ & $1$            \\ 
                               &  2 &  3 & $( 0,  0,  0)$ & $1$            \\ 
                               &  1 &  2 & $( 0, -1,  0)$ & $1$            \\ \hline
\multirow{3}{*}{$J_{3}$  }     &  2 &  1 & $( 0,  0,  1)$ & $\matr{R}_0$   \\ 
                               &  3 &  2 & $( 1,  1,  1)$ & $\matr{R}_0$   \\ 
                               &  1 &  3 & $(-1, -1,  1)$ & $\matr{R}_0$   \\ \hline
\multirow{3}{*}{$J_{4}$  }     &  1 &  2 & $( 0,  0,  1)$ & $\matr{R}_0$   \\ 
                               &  2 &  2 & $( 0,  0,  1)$ & $\matr{R}_0$   \\ 
                               &  3 &  2 & $( 0,  0,  1)$ & $\matr{R}_0$   \\ \hline
\multirow{3}{*}{$J_{5}$  }     &  1 &  2 & $( 0,  0,  1)$ & $\matr{R}_0$   \\ 
                               &  2 &  3 & $(-1, -1,  1)$ & $\matr{R}_0$   \\ 
                               &  3 &  1 & $( 1,  1,  1)$ & $\matr{R}_0$   \\ \hline
        		\end{tabular}
\end{table}
\end{ruledtabular}

\section{Summary}\label{sSum}

In this paper we described a general algorithm to calculate dynamical spin-spin correlation function using linear spin wave theory on magnetic lattices with incommensurate order. The method can accommodate models where the interacting atoms have different spin quantum numbers. It also includes a general single ion anisotropy, anisotropic and antisymmetric exchange interactions. The main idea behind the general solution is to define a local coordinate system that transforms the incommensurate magnetic structure into ferromagnetic order by the consecutive application of two rotations. First a global rotation, that transforms the incommensurate structure into a commensurate one. Secondly a local rotation on every moment within the crystallographic unit cells. This method enables the calculation of the spin wave spectrum of incommensurate systems, that was achieved without using a large supercell and an approximate ordering wave vector \cite{Damay2011,Loire2011,Poienar2010}. We also showed the necessary steps that define an algorithm, that is available under GNU general public license \cite{Toth2013}. Finally as an example we showed how the algorithm can be used to calculate the spectrum of magnetic C-site langasites with incommensurate magnetic order \cite{Marty2010}.

\begin{acknowledgments}
The research leading to these results has received funding from the European Community's Seventh Framework Programme (FP7/2007-2013) under grant agreement n.$^\circ$ 290605  (COFUND: PSI-FELLOW).
\end{acknowledgments}

%\bibliography{library}

\begin{thebibliography}{33}%
\makeatletter
\providecommand \@ifxundefined [1]{%
 \@ifx{#1\undefined}
}%
\providecommand \@ifnum [1]{%
 \ifnum #1\expandafter \@firstoftwo
 \else \expandafter \@secondoftwo
 \fi
}%
\providecommand \@ifx [1]{%
 \ifx #1\expandafter \@firstoftwo
 \else \expandafter \@secondoftwo
 \fi
}%
\providecommand \natexlab [1]{#1}%
\providecommand \enquote  [1]{``#1''}%
\providecommand \bibnamefont  [1]{#1}%
\providecommand \bibfnamefont [1]{#1}%
\providecommand \citenamefont [1]{#1}%
\providecommand \href@noop [0]{\@secondoftwo}%
\providecommand \href [0]{\begingroup \@sanitize@url \@href}%
\providecommand \@href[1]{\@@startlink{#1}\@@href}%
\providecommand \@@href[1]{\endgroup#1\@@endlink}%
\providecommand \@sanitize@url [0]{\catcode `\\12\catcode `\$12\catcode
  `\&12\catcode `\#12\catcode `\^12\catcode `\_12\catcode `\%12\relax}%
\providecommand \@@startlink[1]{}%
\providecommand \@@endlink[0]{}%
\providecommand \url  [0]{\begingroup\@sanitize@url \@url }%
\providecommand \@url [1]{\endgroup\@href {#1}{\urlprefix }}%
\providecommand \urlprefix  [0]{URL }%
\providecommand \Eprint [0]{\href }%
\providecommand \doibase [0]{http://dx.doi.org/}%
\providecommand \selectlanguage [0]{\@gobble}%
\providecommand \bibinfo  [0]{\@secondoftwo}%
\providecommand \bibfield  [0]{\@secondoftwo}%
\providecommand \translation [1]{[#1]}%
\providecommand \BibitemOpen [0]{}%
\providecommand \bibitemStop [0]{}%
\providecommand \bibitemNoStop [0]{.\EOS\space}%
\providecommand \EOS [0]{\spacefactor3000\relax}%
\providecommand \BibitemShut  [1]{\csname bibitem#1\endcsname}%
\let\auto@bib@innerbib\@empty
%</preamble>
\bibitem [{\citenamefont {Bloch}(1930)}]{Bloch1930a}%
  \BibitemOpen
  \bibfield  {author} {\bibinfo {author} {\bibfnamefont {F.}~\bibnamefont
  {Bloch}},\ }\href {\doibase 10.1007/BF01339661} {\bibfield  {journal}
  {\bibinfo  {journal} {Zeitschrift f\"{u}r Phys.}\ }\textbf {\bibinfo {volume}
  {61}},\ \bibinfo {pages} {206} (\bibinfo {year} {1930})}\BibitemShut
  {NoStop}%
\bibitem [{\citenamefont {Slater}(1930)}]{Slater1930}%
  \BibitemOpen
  \bibfield  {author} {\bibinfo {author} {\bibfnamefont {J.~C.}\ \bibnamefont
  {Slater}},\ }\href {\doibase 10.1103/PhysRev.35.509} {\bibfield  {journal}
  {\bibinfo  {journal} {Phys. Rev.}\ }\textbf {\bibinfo {volume} {35}},\
  \bibinfo {pages} {509} (\bibinfo {year} {1930})}\BibitemShut {NoStop}%
\bibitem [{\citenamefont {Holstein}\ and\ \citenamefont
  {Primakoff}(1940)}]{Holstein1940a}%
  \BibitemOpen
  \bibfield  {author} {\bibinfo {author} {\bibfnamefont {T.}~\bibnamefont
  {Holstein}}\ and\ \bibinfo {author} {\bibfnamefont {H.}~\bibnamefont
  {Primakoff}},\ }\href {\doibase 10.1103/PhysRev.58.1098} {\bibfield
  {journal} {\bibinfo  {journal} {Phys. Rev.}\ }\textbf {\bibinfo {volume}
  {58}},\ \bibinfo {pages} {1098} (\bibinfo {year} {1940})}\BibitemShut
  {NoStop}%
\bibitem [{\citenamefont {Dyson}(1956{\natexlab{a}})}]{Dyson1956b}%
  \BibitemOpen
  \bibfield  {author} {\bibinfo {author} {\bibfnamefont {F.}~\bibnamefont
  {Dyson}},\ }\href {\doibase 10.1103/PhysRev.102.1230} {\bibfield  {journal}
  {\bibinfo  {journal} {Phys. Rev.}\ }\textbf {\bibinfo {volume} {102}},\
  \bibinfo {pages} {1230} (\bibinfo {year} {1956}{\natexlab{a}})}\BibitemShut
  {NoStop}%
\bibitem [{\citenamefont {Dyson}(1956{\natexlab{b}})}]{Dyson1956}%
  \BibitemOpen
  \bibfield  {author} {\bibinfo {author} {\bibfnamefont {F.~J.}\ \bibnamefont
  {Dyson}},\ }\href {\doibase 10.1103/PhysRev.102.1217} {\bibfield  {journal}
  {\bibinfo  {journal} {Phys. Rev.}\ }\textbf {\bibinfo {volume} {102}},\
  \bibinfo {pages} {1217} (\bibinfo {year} {1956}{\natexlab{b}})}\BibitemShut
  {NoStop}%
\bibitem [{\citenamefont {Fennell}\ \emph {et~al.}(2009)\citenamefont
  {Fennell}, \citenamefont {Deen}, \citenamefont {Wildes}, \citenamefont
  {Schmalzl}, \citenamefont {Prabhakaran}, \citenamefont {Boothroyd},
  \citenamefont {Aldus}, \citenamefont {McMorrow},\ and\ \citenamefont
  {Bramwell}}]{Fennell2009}%
  \BibitemOpen
  \bibfield  {author} {\bibinfo {author} {\bibfnamefont {T.}~\bibnamefont
  {Fennell}}, \bibinfo {author} {\bibfnamefont {P.~P.}\ \bibnamefont {Deen}},
  \bibinfo {author} {\bibfnamefont {A.~R.}\ \bibnamefont {Wildes}}, \bibinfo
  {author} {\bibfnamefont {K.}~\bibnamefont {Schmalzl}}, \bibinfo {author}
  {\bibfnamefont {D.}~\bibnamefont {Prabhakaran}}, \bibinfo {author}
  {\bibfnamefont {A.~T.}\ \bibnamefont {Boothroyd}}, \bibinfo {author}
  {\bibfnamefont {R.~J.}\ \bibnamefont {Aldus}}, \bibinfo {author}
  {\bibfnamefont {D.~F.}\ \bibnamefont {McMorrow}}, \ and\ \bibinfo {author}
  {\bibfnamefont {S.~T.}\ \bibnamefont {Bramwell}},\ }\href {\doibase
  10.1126/science.1177582} {\bibfield  {journal} {\bibinfo  {journal} {Science}
  }\textbf {\bibinfo {volume} {326}},\ \bibinfo {pages} {415}
  (\bibinfo {year} {2009})}\BibitemShut {NoStop}%
\bibitem [{\citenamefont {Morris}\ \emph {et~al.}(2009)\citenamefont {Morris},
  \citenamefont {Tennant}, \citenamefont {Grigera}, \citenamefont {Klemke},
  \citenamefont {Castelnovo}, \citenamefont {Moessner}, \citenamefont
  {Czternasty}, \citenamefont {Meissner}, \citenamefont {Rule}, \citenamefont
  {Hoffmann}, \citenamefont {Kiefer}, \citenamefont {Gerischer}, \citenamefont
  {Slobinsky},\ and\ \citenamefont {Perry}}]{Morris2009}%
  \BibitemOpen
  \bibfield  {author} {\bibinfo {author} {\bibfnamefont {D.~J.~P.}\
  \bibnamefont {Morris}}, \bibinfo {author} {\bibfnamefont {D.~A.}\
  \bibnamefont {Tennant}}, \bibinfo {author} {\bibfnamefont {S.~A.}\
  \bibnamefont {Grigera}}, \bibinfo {author} {\bibfnamefont {B.}~\bibnamefont
  {Klemke}}, \bibinfo {author} {\bibfnamefont {C.}~\bibnamefont {Castelnovo}},
  \bibinfo {author} {\bibfnamefont {R.}~\bibnamefont {Moessner}}, \bibinfo
  {author} {\bibfnamefont {C.}~\bibnamefont {Czternasty}}, \bibinfo {author}
  {\bibfnamefont {M.}~\bibnamefont {Meissner}}, \bibinfo {author}
  {\bibfnamefont {K.~C.}\ \bibnamefont {Rule}}, \bibinfo {author}
  {\bibfnamefont {J.-U.}\ \bibnamefont {Hoffmann}}, \bibinfo {author}
  {\bibfnamefont {K.}~\bibnamefont {Kiefer}}, \bibinfo {author} {\bibfnamefont
  {S.}~\bibnamefont {Gerischer}}, \bibinfo {author} {\bibfnamefont
  {D.}~\bibnamefont {Slobinsky}}, \ and\ \bibinfo {author} {\bibfnamefont
  {R.~S.}\ \bibnamefont {Perry}},\ }\href {\doibase 10.1126/science.1178868}
  {\bibfield  {journal} {\bibinfo  {journal} {Science}\ }\textbf
  {\bibinfo {volume} {326}},\ \bibinfo {pages} {411} (\bibinfo {year}
  {2009})}\BibitemShut {NoStop}%
\bibitem [{\citenamefont {Balents}(2010)}]{Balents2010}%
  \BibitemOpen
  \bibfield  {author} {\bibinfo {author} {\bibfnamefont {L.}~\bibnamefont
  {Balents}},\ }\href {\doibase 10.1038/nature08917} {\bibfield  {journal}
  {\bibinfo  {journal} {Nature}\ }\textbf {\bibinfo {volume} {464}},\ \bibinfo
  {pages} {199} (\bibinfo {year} {2010})}\BibitemShut {NoStop}%
\bibitem [{\citenamefont {Khomskii}(2009)}]{Khomskii2009a}%
  \BibitemOpen
  \bibfield  {author} {\bibinfo {author} {\bibfnamefont {D.}~\bibnamefont
  {Khomskii}},\ }\href {\doibase 10.1103/Physics.2.20} {\bibfield  {journal}
  {\bibinfo  {journal} {Physics}\ }\textbf {\bibinfo
  {volume} {2}},\ \bibinfo {pages} {20} (\bibinfo {year} {2009})}\BibitemShut
  {NoStop}%
\bibitem [{\citenamefont {Kim}\ \emph {et~al.}(2012)\citenamefont {Kim},
  \citenamefont {Said}, \citenamefont {Casa}, \citenamefont {Upton},
  \citenamefont {Gog}, \citenamefont {Daghofer}, \citenamefont {Jackeli},
  \citenamefont {van~den Brink}, \citenamefont {Khaliullin},\ and\
  \citenamefont {Kim}}]{Kim2012a}%
  \BibitemOpen
  \bibfield  {author} {\bibinfo {author} {\bibfnamefont {J.}~\bibnamefont
  {Kim}}, \bibinfo {author} {\bibfnamefont {a.~H.}\ \bibnamefont {Said}},
  \bibinfo {author} {\bibfnamefont {D.}~\bibnamefont {Casa}}, \bibinfo {author}
  {\bibfnamefont {M.~H.}\ \bibnamefont {Upton}}, \bibinfo {author}
  {\bibfnamefont {T.}~\bibnamefont {Gog}}, \bibinfo {author} {\bibfnamefont
  {M.}~\bibnamefont {Daghofer}}, \bibinfo {author} {\bibfnamefont
  {G.}~\bibnamefont {Jackeli}}, \bibinfo {author} {\bibfnamefont
  {J.}~\bibnamefont {van~den Brink}}, \bibinfo {author} {\bibfnamefont
  {G.}~\bibnamefont {Khaliullin}}, \ and\ \bibinfo {author} {\bibfnamefont
  {B.~J.}\ \bibnamefont {Kim}},\ }\href {\doibase
  10.1103/PhysRevLett.109.157402} {\bibfield  {journal} {\bibinfo  {journal}
  {Phys. Rev. Lett.}\ }\textbf {\bibinfo {volume} {109}},\ \bibinfo {pages}
  {157402} (\bibinfo {year} {2012})}\BibitemShut {NoStop}%
\bibitem [{\citenamefont {Petit}(2011)}]{Petit2011}%
  \BibitemOpen
  \bibfield  {author} {\bibinfo {author} {\bibfnamefont {S.}~\bibnamefont
  {Petit}},\ }\href {\doibase 10.1051/sfn/201112006} {\bibfield  {journal}
  {\bibinfo  {journal} {Collection SFN}\ }\textbf {\bibinfo {volume} {12}},\ \bibinfo
  {pages} {105} (\bibinfo {year} {2011})}\BibitemShut {NoStop}%
\bibitem [{\citenamefont {Haraldsen}(2010)}]{Haraldsen2010b}%
  \BibitemOpen
  \bibfield  {author} {\bibinfo {author} {\bibfnamefont {J.}~\bibnamefont
  {Haraldsen}}, \ and\ \bibinfo {author} {\bibfnamefont
  {R.}\ \bibnamefont {Fishman}},\ }  \href {\doibase 10.1088/0953-8984/22/50/509801} {\bibfield  {journal}
  {\bibinfo  {journal} {J. Phys. Cond. Mat.}\ }\textbf {\bibinfo {volume} {21}},\ \bibinfo
  {pages} {216001} (\bibinfo {year} {2009})}\BibitemShut {NoStop}%
\bibitem [{\citenamefont {Zhitomirsky}\ and\ \citenamefont
  {Chernyshev}(2013)}]{Zhitomirsky2012}%
  \BibitemOpen
  \bibfield  {author} {\bibinfo {author} {\bibfnamefont {M.~E.}\ \bibnamefont
  {Zhitomirsky}}\ and\ \bibinfo {author} {\bibfnamefont {a.~L.}\ \bibnamefont
  {Chernyshev}},\ }\href {\doibase 10.1103/RevModPhys.85.219} {\bibfield
  {journal} {\bibinfo  {journal} {Rev. Mod. Phys.}\ }\textbf {\bibinfo {volume}
  {85}},\ \bibinfo {pages} {219} (\bibinfo {year} {2013})}\BibitemShut {NoStop}%
\bibitem [{\citenamefont {Mourigal}\ \emph {et~al.}(2013)\citenamefont
  {Mourigal}, \citenamefont {Fuhrman}, \citenamefont {Chernyshev},\ and\
  \citenamefont {Zhitomirsky}}]{Mourigal2013}%
  \BibitemOpen
  \bibfield  {author} {\bibinfo {author} {\bibfnamefont {M.}~\bibnamefont
  {Mourigal}}, \bibinfo {author} {\bibfnamefont {W.~T.}\ \bibnamefont
  {Fuhrman}}, \bibinfo {author} {\bibfnamefont {a.~L.}\ \bibnamefont
  {Chernyshev}}, \ and\ \bibinfo {author} {\bibfnamefont {M.~E.}\ \bibnamefont
  {Zhitomirsky}},\ }\href {\doibase 10.1103/PhysRevB.88.094407} {\bibfield
  {journal} {\bibinfo  {journal} {Phys. Rev. B}\ }\textbf {\bibinfo {volume}
  {88}},\ \bibinfo {pages} {094407} (\bibinfo {year} {2013})}\BibitemShut
  {NoStop}%
\bibitem [{\citenamefont {Toth}(2014)}]{Toth2013}%
  \BibitemOpen
  \bibfield  {author} {\bibinfo {author} {\bibfnamefont {S.}~\bibnamefont
  {Toth}},\ }\url {http://www.psi.ch/spinw} (\bibinfo {year} {2014})\BibitemShut {NoStop}%
\bibitem [{\citenamefont {Colpa}(1978)}]{Colpa1978}%
  \BibitemOpen
  \bibfield  {author} {\bibinfo {author} {\bibfnamefont {J.}~\bibnamefont
  {Colpa}},\ }\href {\doibase 10.1016/0378-4371(78)90160-7} {\bibfield
  {journal} {\bibinfo  {journal} {Phys. A Stat. Mech. its Appl.}\ }\textbf
  {\bibinfo {volume} {93}},\ \bibinfo {pages} {327} (\bibinfo {year}
  {1978})}\BibitemShut {NoStop}%
\bibitem [{\citenamefont {Kaplan}(1961)}]{Kaplan1961}%
  \BibitemOpen
  \bibfield  {author} {\bibinfo {author} {\bibfnamefont {T.}~\bibnamefont
  {Kaplan}},\ }\href {\doibase 10.1103/PhysRev.124.329} {\bibfield  {journal}
  {\bibinfo  {journal} {Phys. Rev.}\ }\textbf {\bibinfo {volume} {124}},\
  \bibinfo {pages} {329} (\bibinfo {year} {1961})}\BibitemShut {NoStop}%
\bibitem [{\citenamefont {Chernyshev}\ and\ \citenamefont
  {Zhitomirsky}(2009)}]{Chernyshev2009}%
  \BibitemOpen
  \bibfield  {author} {\bibinfo {author} {\bibfnamefont {A.}~\bibnamefont
  {Chernyshev}}\ and\ \bibinfo {author} {\bibfnamefont {M.~E.}\ \bibnamefont
  {Zhitomirsky}},\ }\href {\doibase 10.1103/PhysRevB.79.144416} {\bibfield
  {journal} {\bibinfo  {journal} {Phys. Rev. B}\ }\textbf {\bibinfo {volume}
  {79}},\ \bibinfo {pages} {144416} (\bibinfo {year} {2009})}\BibitemShut
  {NoStop}%
\bibitem [{\citenamefont {Damay}\ \emph {et~al.}(2010)\citenamefont {Damay},
  \citenamefont {Martin}, \citenamefont {Hardy}, \citenamefont {Maignan},
  \citenamefont {Andr\'{e}}, \citenamefont {Knight}, \citenamefont {Giblin},\
  and\ \citenamefont {Chapon}}]{Damay2010}%
  \BibitemOpen
  \bibfield  {author} {\bibinfo {author} {\bibfnamefont {F.}~\bibnamefont
  {Damay}}, \bibinfo {author} {\bibfnamefont {C.}~\bibnamefont {Martin}},
  \bibinfo {author} {\bibfnamefont {V.}~\bibnamefont {Hardy}}, \bibinfo
  {author} {\bibfnamefont {A.}~\bibnamefont {Maignan}}, \bibinfo {author}
  {\bibfnamefont {G.}~\bibnamefont {Andr\'{e}}}, \bibinfo {author}
  {\bibfnamefont {K.}~\bibnamefont {Knight}}, \bibinfo {author} {\bibfnamefont
  {S.~R.}\ \bibnamefont {Giblin}}, \ and\ \bibinfo {author} {\bibfnamefont
  {L.~C.}\ \bibnamefont {Chapon}},\ }\href {\doibase
  10.1103/PhysRevB.81.214405} {\bibfield  {journal} {\bibinfo  {journal} {Phys.
  Rev. B}\ }\textbf {\bibinfo {volume} {81}},\ \bibinfo {pages} {214405}
  (\bibinfo {year} {2010})}\BibitemShut {NoStop}%
\bibitem [{\citenamefont {Luttinger}\ and\ \citenamefont
  {Tisza}(1946)}]{Luttinger1946a}%
  \BibitemOpen
  \bibfield  {author} {\bibinfo {author} {\bibfnamefont {J.}~\bibnamefont
  {Luttinger}}\ and\ \bibinfo {author} {\bibfnamefont {L.}~\bibnamefont
  {Tisza}},\ }\href {\doibase 10.1103/PhysRev.70.954} {\bibfield  {journal}
  {\bibinfo  {journal} {Phys. Rev.}\ }\textbf {\bibinfo {volume} {70}},\
  \bibinfo {pages} {954} (\bibinfo {year} {1946})}\BibitemShut {NoStop}%
\bibitem [{\citenamefont {Kaplan}\ and\ \citenamefont
  {Menyuk}(2007)}]{Kaplan2007}%
  \BibitemOpen
  \bibfield  {author} {\bibinfo {author} {\bibfnamefont {T.~a.}\ \bibnamefont
  {Kaplan}}\ and\ \bibinfo {author} {\bibfnamefont {N.}~\bibnamefont
  {Menyuk}},\ }\href {\doibase 10.1080/14786430601080229} {\bibfield  {journal}
  {\bibinfo  {journal} {Philos. Mag.}\ }\textbf {\bibinfo {volume} {87}},\
  \bibinfo {pages} {3711} (\bibinfo {year} {2007})}\BibitemShut {NoStop}%
\bibitem [{\citenamefont {Lacorre}\ and\ \citenamefont
  {Pannetier}(1987)}]{Lacorre1987}%
  \BibitemOpen
  \bibfield  {author} {\bibinfo {author} {\bibfnamefont {P.}~\bibnamefont
  {Lacorre}}\ and\ \bibinfo {author} {\bibfnamefont {J.}~\bibnamefont
  {Pannetier}},\ }\href {\doibase 10.1016/0304-8853(87)90334-9} {\bibfield
  {journal} {\bibinfo  {journal} {J. Magn. Magn. Mater.}\ }\textbf {\bibinfo
  {volume} {71}},\ \bibinfo {pages} {63} (\bibinfo {year} {1987})}\BibitemShut
  {NoStop}%
\bibitem [{\citenamefont {Marshall}\ and\ \citenamefont
  {Lovesey}(1971)}]{Marshall1971}%
  \BibitemOpen
  \bibfield  {author} {\bibinfo {author} {\bibfnamefont {W.}~\bibnamefont
  {Marshall}}\ and\ \bibinfo {author} {\bibfnamefont {S.~W.}\ \bibnamefont
  {Lovesey}}\ }(\bibinfo  {publisher} {Clarendon Press},\ \bibinfo {address}
  {Oxford},\ \bibinfo {year} {1971})\ Chap.~\bibinfo {chapter} {9}\BibitemShut
  {NoStop}%
\bibitem [{\citenamefont {Coldea}\ \emph {et~al.}(2003)\citenamefont {Coldea},
  \citenamefont {Tennant},\ and\ \citenamefont {Tylczynski}}]{Coldea2003}%
  \BibitemOpen
  \bibfield  {author} {\bibinfo {author} {\bibfnamefont {R.}~\bibnamefont
  {Coldea}}, \bibinfo {author} {\bibfnamefont {D.}~\bibnamefont {Tennant}}, \
  and\ \bibinfo {author} {\bibfnamefont {Z.}~\bibnamefont {Tylczynski}},\
  }\href {\doibase 10.1103/PhysRevB.68.134424} {\bibfield  {journal} {\bibinfo
  {journal} {Phys. Rev. B}\ }\textbf {\bibinfo {volume} {68}},\ \bibinfo
  {pages} {134424} (\bibinfo {year} {2003})}\BibitemShut {NoStop}%
\bibitem [{\citenamefont {Iwataki}\ \emph {et~al.}(2001)\citenamefont
  {Iwataki}, \citenamefont {Ohsato}, \citenamefont {Tanaka}, \citenamefont
  {Morikoshi}, \citenamefont {Sato},\ and\ \citenamefont
  {Kawasaki}}]{Iwataki2001}%
  \BibitemOpen
  \bibfield  {author} {\bibinfo {author} {\bibfnamefont {T.}~\bibnamefont
  {Iwataki}}, \bibinfo {author} {\bibfnamefont {H.}~\bibnamefont {Ohsato}},
  \bibinfo {author} {\bibfnamefont {K.}~\bibnamefont {Tanaka}}, \bibinfo
  {author} {\bibfnamefont {H.}~\bibnamefont {Morikoshi}}, \bibinfo {author}
  {\bibfnamefont {J.}~\bibnamefont {Sato}}, \ and\ \bibinfo {author}
  {\bibfnamefont {K.}~\bibnamefont {Kawasaki}},\ }\href {\doibase
  10.1016/S0955-2219(01)00029-2} {\bibfield  {journal} {\bibinfo  {journal} {J.
  Eur. Ceram. Soc.}\ }\textbf {\bibinfo {volume} {21}},\ \bibinfo {pages}
  {1409} (\bibinfo {year} {2001})}\BibitemShut {NoStop}%
\bibitem [{\citenamefont {Fritze}\ and\ \citenamefont
  {Tuller}(2001)}]{Fritze2001}%
  \BibitemOpen
  \bibfield  {author} {\bibinfo {author} {\bibfnamefont {H.}~\bibnamefont
  {Fritze}}\ and\ \bibinfo {author} {\bibfnamefont {H.~L.}\ \bibnamefont
  {Tuller}},\ }\href {\doibase 10.1063/1.1345797} {\bibfield  {journal}
  {\bibinfo  {journal} {Appl. Phys. Lett.}\ }\textbf {\bibinfo {volume} {78}},\
  \bibinfo {pages} {976} (\bibinfo {year} {2001})}\BibitemShut {NoStop}%
\bibitem [{\citenamefont {Bordet}\ \emph {et~al.}(2006)\citenamefont {Bordet},
  \citenamefont {Gelard}, \citenamefont {Marty}, \citenamefont {Ibanez},
  \citenamefont {Robert}, \citenamefont {Simonet}, \citenamefont {Canals},
  \citenamefont {Ballou},\ and\ \citenamefont {Lejay}}]{Bordet2006}%
  \BibitemOpen
  \bibfield  {author} {\bibinfo {author} {\bibfnamefont {P.}~\bibnamefont
  {Bordet}}, \bibinfo {author} {\bibfnamefont {I.}~\bibnamefont {Gelard}},
  \bibinfo {author} {\bibfnamefont {K.}~\bibnamefont {Marty}}, \bibinfo
  {author} {\bibfnamefont {A.}~\bibnamefont {Ibanez}}, \bibinfo {author}
  {\bibfnamefont {J.}~\bibnamefont {Robert}}, \bibinfo {author} {\bibfnamefont
  {V.}~\bibnamefont {Simonet}}, \bibinfo {author} {\bibfnamefont
  {B.}~\bibnamefont {Canals}}, \bibinfo {author} {\bibfnamefont
  {R.}~\bibnamefont {Ballou}}, \ and\ \bibinfo {author} {\bibfnamefont
  {P.}~\bibnamefont {Lejay}},\ }\href {\doibase 10.1088/0953-8984/18/22/014}
  {\bibfield  {journal} {\bibinfo  {journal} {J. Phys. Condens. Matter}\
  }\textbf {\bibinfo {volume} {18}},\ \bibinfo {pages} {5147} (\bibinfo {year}
  {2006})}\BibitemShut {NoStop}%
\bibitem [{\citenamefont {Marty}\ \emph {et~al.}(2008)\citenamefont {Marty},
  \citenamefont {Simonet}, \citenamefont {Ressouche}, \citenamefont {Ballou},
  \citenamefont {Lejay},\ and\ \citenamefont {Bordet}}]{Marty2008}%
  \BibitemOpen
  \bibfield  {author} {\bibinfo {author} {\bibfnamefont {K.}~\bibnamefont
  {Marty}}, \bibinfo {author} {\bibfnamefont {V.}~\bibnamefont {Simonet}},
  \bibinfo {author} {\bibfnamefont {E.}~\bibnamefont {Ressouche}}, \bibinfo
  {author} {\bibfnamefont {R.}~\bibnamefont {Ballou}}, \bibinfo {author}
  {\bibfnamefont {P.}~\bibnamefont {Lejay}}, \ and\ \bibinfo {author}
  {\bibfnamefont {P.}~\bibnamefont {Bordet}},\ }\href {\doibase
  10.1103/PhysRevLett.101.247201} {\bibfield  {journal} {\bibinfo  {journal}
  {Phys. Rev. Lett.}\ }\textbf {\bibinfo {volume} {101}},\ \bibinfo {pages}
  {247201} (\bibinfo {year} {2008})}\BibitemShut {NoStop}%
\bibitem [{\citenamefont {Marty}\ \emph {et~al.}(2010)\citenamefont {Marty},
  \citenamefont {Bordet}, \citenamefont {Simonet}, \citenamefont {Loire},
  \citenamefont {Ballou}, \citenamefont {Darie}, \citenamefont {Kljun},
  \citenamefont {Bonville}, \citenamefont {Isnard}, \citenamefont {Lejay},
  \citenamefont {Zawilski},\ and\ \citenamefont {Simon}}]{Marty2010}%
  \BibitemOpen
  \bibfield  {author} {\bibinfo {author} {\bibfnamefont {K.}~\bibnamefont
  {Marty}}, \bibinfo {author} {\bibfnamefont {P.}~\bibnamefont {Bordet}},
  \bibinfo {author} {\bibfnamefont {V.}~\bibnamefont {Simonet}}, \bibinfo
  {author} {\bibfnamefont {M.}~\bibnamefont {Loire}}, \bibinfo {author}
  {\bibfnamefont {R.}~\bibnamefont {Ballou}}, \bibinfo {author} {\bibfnamefont
  {C.}~\bibnamefont {Darie}}, \bibinfo {author} {\bibfnamefont
  {J.}~\bibnamefont {Kljun}}, \bibinfo {author} {\bibfnamefont
  {P.}~\bibnamefont {Bonville}}, \bibinfo {author} {\bibfnamefont
  {O.}~\bibnamefont {Isnard}}, \bibinfo {author} {\bibfnamefont
  {P.}~\bibnamefont {Lejay}}, \bibinfo {author} {\bibfnamefont
  {B.}~\bibnamefont {Zawilski}}, \ and\ \bibinfo {author} {\bibfnamefont
  {C.}~\bibnamefont {Simon}},\ }\href {\doibase 10.1103/PhysRevB.81.054416}
  {\bibfield  {journal} {\bibinfo  {journal} {Phys. Rev. B}\ }\textbf {\bibinfo
  {volume} {81}},\ \bibinfo {pages} {054416} (\bibinfo {year}
  {2010})}\BibitemShut {NoStop}%
\bibitem [{\citenamefont {Stock}\ \emph {et~al.}(2011)\citenamefont {Stock},
  \citenamefont {Chapon}, \citenamefont {Schneidewind}, \citenamefont {Su},
  \citenamefont {Radaelli}, \citenamefont {McMorrow}, \citenamefont {Bombardi},
  \citenamefont {Lee},\ and\ \citenamefont {Cheong}}]{Stock2011}%
  \BibitemOpen
  \bibfield  {author} {\bibinfo {author} {\bibfnamefont {C.}~\bibnamefont
  {Stock}}, \bibinfo {author} {\bibfnamefont {L.~C.}\ \bibnamefont {Chapon}},
  \bibinfo {author} {\bibfnamefont {A.}~\bibnamefont {Schneidewind}}, \bibinfo
  {author} {\bibfnamefont {Y.}~\bibnamefont {Su}}, \bibinfo {author}
  {\bibfnamefont {P.~G.}\ \bibnamefont {Radaelli}}, \bibinfo {author}
  {\bibfnamefont {D.~F.}\ \bibnamefont {McMorrow}}, \bibinfo {author}
  {\bibfnamefont {A.}~\bibnamefont {Bombardi}}, \bibinfo {author}
  {\bibfnamefont {N.}~\bibnamefont {Lee}}, \ and\ \bibinfo {author}
  {\bibfnamefont {S.-W.}\ \bibnamefont {Cheong}},\ }\href {\doibase
  10.1103/PhysRevB.83.104426} {\bibfield  {journal} {\bibinfo  {journal} {Phys.
  Rev. B}\ }\textbf {\bibinfo {volume} {83}},\ \bibinfo {pages} {104426}
  (\bibinfo {year} {2011})}\BibitemShut {NoStop}%
\bibitem [{\citenamefont {Loire}\ \emph {et~al.}(2011)\citenamefont {Loire},
  \citenamefont {Simonet}, \citenamefont {Petit}, \citenamefont {Marty},
  \citenamefont {Bordet}, \citenamefont {Lejay}, \citenamefont {Ollivier},
  \citenamefont {Enderle}, \citenamefont {Steffens}, \citenamefont {Ressouche},
  \citenamefont {Zorko},\ and\ \citenamefont {Ballou}}]{Loire2011}%
  \BibitemOpen
  \bibfield  {author} {\bibinfo {author} {\bibfnamefont {M.}~\bibnamefont
  {Loire}}, \bibinfo {author} {\bibfnamefont {V.}~\bibnamefont {Simonet}},
  \bibinfo {author} {\bibfnamefont {S.}~\bibnamefont {Petit}}, \bibinfo
  {author} {\bibfnamefont {K.}~\bibnamefont {Marty}}, \bibinfo {author}
  {\bibfnamefont {P.}~\bibnamefont {Bordet}}, \bibinfo {author} {\bibfnamefont
  {P.}~\bibnamefont {Lejay}}, \bibinfo {author} {\bibfnamefont
  {J.}~\bibnamefont {Ollivier}}, \bibinfo {author} {\bibfnamefont
  {M.}~\bibnamefont {Enderle}}, \bibinfo {author} {\bibfnamefont
  {P.}~\bibnamefont {Steffens}}, \bibinfo {author} {\bibfnamefont
  {E.}~\bibnamefont {Ressouche}}, \bibinfo {author} {\bibfnamefont
  {A.}~\bibnamefont {Zorko}}, \ and\ \bibinfo {author} {\bibfnamefont
  {R.}~\bibnamefont {Ballou}},\ }\href {\doibase
  10.1103/PhysRevLett.106.207201} {\bibfield  {journal} {\bibinfo  {journal}
  {Phys. Rev. Lett.}\ }\textbf {\bibinfo {volume} {106}},\ \bibinfo {pages}
  {207201} (\bibinfo {year} {2011})}\BibitemShut {NoStop}%
\bibitem [{\citenamefont {Jensen}(2011)}]{Jensen2011}%
  \BibitemOpen
  \bibfield  {author} {\bibinfo {author} {\bibfnamefont {J.}~\bibnamefont
  {Jensen}},\ }\href {\doibase 10.1103/PhysRevB.84.104405} {\bibfield
  {journal} {\bibinfo  {journal} {Phys. Rev. B}\ }\textbf {\bibinfo {volume}
  {84}},\ \bibinfo {pages} {104405} (\bibinfo {year} {2011})}\BibitemShut
  {NoStop}%
\bibitem [{\citenamefont {Damay}\ \emph {et~al.}(2011)\citenamefont {Damay},
  \citenamefont {Martin}, \citenamefont {Hardy}, \citenamefont {Maignan},
  \citenamefont {Stock},\ and\ \citenamefont {Petit}}]{Damay2011}%
  \BibitemOpen
  \bibfield  {author} {\bibinfo {author} {\bibfnamefont {F.}~\bibnamefont
  {Damay}}, \bibinfo {author} {\bibfnamefont {C.}~\bibnamefont {Martin}},
  \bibinfo {author} {\bibfnamefont {V.}~\bibnamefont {Hardy}}, \bibinfo
  {author} {\bibfnamefont {A.}~\bibnamefont {Maignan}}, \bibinfo {author}
  {\bibfnamefont {C.}~\bibnamefont {Stock}}, \ and\ \bibinfo {author}
  {\bibfnamefont {S.}~\bibnamefont {Petit}},\ }\href {\doibase
  10.1103/PhysRevB.84.020402} {\bibfield  {journal} {\bibinfo  {journal} {Phys.
  Rev. B}\ }\textbf {\bibinfo {volume} {84}},\ \bibinfo {pages} {020402}
  (\bibinfo {year} {2011})}\BibitemShut {NoStop}%
\bibitem [{\citenamefont {Poienar}\ \emph {et~al.}(2010)\citenamefont
  {Poienar}, \citenamefont {Damay}, \citenamefont {Martin}, \citenamefont
  {Robert},\ and\ \citenamefont {Petit}}]{Poienar2010}%
  \BibitemOpen
  \bibfield  {author} {\bibinfo {author} {\bibfnamefont {M.}~\bibnamefont
  {Poienar}}, \bibinfo {author} {\bibfnamefont {F.}~\bibnamefont {Damay}},
  \bibinfo {author} {\bibfnamefont {C.}~\bibnamefont {Martin}}, \bibinfo
  {author} {\bibfnamefont {J.}~\bibnamefont {Robert}}, \ and\ \bibinfo {author}
  {\bibfnamefont {S.}~\bibnamefont {Petit}},\ }\href {\doibase
  10.1103/PhysRevB.81.104411} {\bibfield  {journal} {\bibinfo  {journal} {Phys.
  Rev. B}\ }\textbf {\bibinfo {volume} {81}},\ \bibinfo {pages} {104411}
  (\bibinfo {year} {2010})}\BibitemShut {NoStop}%
\end{thebibliography}

%merlin.mbs apsrev4-1.bst 2010-07-25 4.21a (PWD, AO, DPC) hacked
%Control: key (0)
%Control: author (8) initials jnrlst
%Control: editor formatted (1) identically to author
%Control: production of article title (-1) disabled
%Control: page (0) single
%Control: year (1) truncated
%Control: production of eprint (0) enabled
%

\end{document}